\documentclass[aps,prd,onecolumn,showpacs,amsmath,amssymb,nofootinbib]{revtex4-1}

\usepackage{epsfig}
\usepackage{graphicx}
\usepackage{dcolumn}
\usepackage{bm}
\usepackage{ltablex,booktabs}
\usepackage{overpic}
\usepackage{subfigure}
\usepackage{float}
\usepackage{color}
\usepackage{amsmath}
\usepackage{mathcomp}
\usepackage{mathrsfs}
\usepackage{multirow}
\usepackage{rotating}
\usepackage{amssymb}
\usepackage{gensymb}
\usepackage{amsmath}
\usepackage{tabularx}

\usepackage[bookmarksnumbered, pdfstartview=FitH,colorlinks,urlcolor=blue, citecolor=blue,linkcolor=blue] {hyperref}

\begin{document}
\parskip=\baselineskip 

\title{Recent Progress in Leptonic and Semileptonic Decays of Charmed Hadrons}
\author{Bai-Cian Ke}
\email[]{baiciank@ihep.ac.cn}
\affiliation{Zhengzhou University, Zhengzhou 450001, People's Republic of China}
\author{Jonna Koponen}
\email[]{jkoponen@uni-mainz.de}
\affiliation{PRISMA+ Cluster of Excellence \& Institute for Nuclear Physics, Johannes Gutenberg University Mainz, Mainz, Germany, D-55128}
\author{Hai-Bo Li}
\email[]{lihb@ihep.ac.cn}
\affiliation{Institute of High Energy Physics, Beijing 100049, People's Republic of China}
\affiliation{University of Chinese Academy of Sciences, Beijing 100049, People's Republic of China}
\author{Yangheng Zheng}
\email[]{zhengyh@ucas.edu.cn}
\affiliation{University of Chinese Academy of Sciences, Beijing 100049, People's Republic of China}


\begin{abstract}
We present a comprehensive review of purely leptonic and semileptonic decays of 
$D^{0(+)}$, $D_s^{+}$, and charmed baryons (including $\Lambda_c^{+}$, $\Xi_c$
and $\Omega_c$). The precise studies of these decays help deepen our
understanding and knowledge of quantum chromodynamics via measuring decay 
constants and form factors, and test the Standard Model through examining 
the unitarity of Cabibbo-Kobayashi-Maskawa matrix and lepton flavor 
universality. We give an overview of the theoretical and experimental tools
before discussing the recent progress. The data sets collected by BESIII near 
the production thresholds of $D\bar{D}$, $D_s^{(*)+}D_s^{(*)-}$ and 
$\Lambda_c^{+}\bar{\Lambda}_c^{-}$ offer important opportunities for studies
of charm physics.
\end{abstract}

\maketitle

\tableofcontents

\section{Introduction}

The Standard Model (SM) is generally accepted as the theory of elementary
particle interactions, with great success in describing a wide variety of
experimental data with energies ranging from under a GeV up to a few
TeV~\cite{PDG}. Quantum chromodynamics (QCD), the theory describing strong
interaction in the SM, has been widely verified by experiments in the high
energy region with its perturbation property, such as ATLAS, CMS, and LHCb
experiments at the Large Hadron Collider (LHC). However, in the low energy
region, the perturbative QCD is not applicable and faces huge challenges from
both theory and experiment. The intriguing color nature of QCD also allows for
the existence of gluon-rich matter, such as hybrids and glueballs, of which
the unambiguous identification and understanding is clearly missing to date.

Furthermore, there are a number of unaddressed issues in the SM, which fails to
explain many phenomena of nature. For example, a yet undiscovered source of
$CP$ violation in the quark sector could explain why matter dominates
antimatter in the universe. We know from astrophysical observations that dark
matter makes up most of the matter in the universe, and the existence of such a
particle would have implications for the SM. Many of the BSM (beyond the SM)
models addressing these questions could affect quark flavor interactions.
Therefore, testing the SM is crucial, and the charm sector provides an
excellent environment to do these tests.

Electron-positron $e^+e^-$ colliders operating at the transition interval
between non-perturbative QCD and perturbative QCD at a few GeV are uniquely
well suited to play an important role in the understanding of strong
interaction and precision test of the SM. Currently, the Beijing
Spectrometer~(BESIII) at the Beijing Electron-Positron Collider~(BEPCII), with
its unique features of threshold characteristics, quantum correlations, etc.,
has addressed a broad physics program covering tests of QCD, investigation of
hadron spectroscopy, precision tests of electroweak interactions, and searches
for new physics in low energy region. 

On the theory side, lattice QCD (LQCD) provides a way to do QCD calculations 
non-perturbatively from first principles. Instead of approximating the QCD 
path integral by expanding it in a series in the strong coupling, $\alpha_s$, 
it can be calculated by discretizing the space-time. This reduces the path
integral to have finite number of degrees of freedom, and imposes a natural
regularization scheme for the theory. Instead of imposing an ultraviolet
cutoff by hand, such as in Pauli-Villars regularization, waves with a
wavelength smaller than $\pi/a$, where $a$ is the lattice spacing, cannot
exist in this discrete space, providing an automatic ultraviolet cutoff.

A typical LQCD simulation uses a space-time grid of $64^3\times 128$ or
$96^3\times 192$, i.e.~the number of degrees of freedom is finite but quite
large. These calculations become possible thanks to the computational power of
modern supercomputers, and the utilization of advanced statistical techniques
and Monte-Carlo integration. Regarding the achievable precision, the challenges
are now different: while in a perturbative calculation the uncertainties come
from the neglected higher order terms, in a LQCD calculation there is the
statistical uncertainty from the Monte Carlo integration as well as systematic
uncertainty from the discretization of the space-time. One needs to introduce a
new dimensionful parameter, the lattice spacing $a$, and one has to extrapolate
to continuum and infinite volume afterwards. The discretization is not unique,
and it usually breaks some symmetries and introduces non-physical particles,
whose contribution has to be suppressed. To some extent, one can choose which
symmetries to preserve (at least partially) by choosing a suitable
discretization scheme. A number of different discretization schemes have been
developed and used over the years, and controlling and estimating the
associated systematic uncertainties has been (and still is) the focus of modern
LQCD calculations. The efforts have been successful, and LQCD has developed
into a precision tool to do QCD calculations non-perturbatively from first 
principles to complement experiments.

In this review article, we give an overview of the theoretical and experimental
tools, before discussing the recent progress in charmed hadron decays. Special
attention is given to new results from BESIII. Charge-conjugated decay modes
are implied throughout this review.

\section{Theoretical framework: CKM matrix elements}
\label{sec:CKM}

The flavor-changing transition of quarks via interaction with a $W^\pm$ boson
is parameterized by the Cabibbo-Kobayashi-Maskawa (CKM) matrix~\cite{CKM}, which 
is a $3\times3$ unitary matrix:
\begin{equation}
V_{\rm CKM}
= \left(
\begin{array}{ccc}
 V_{ud} &  V_{us} &  V_{ub} \\
 V_{cd} &  V_{cs} &  V_{cb} \\
 V_{td} &  V_{ts} &  V_{tb}  \\
\end{array} \right ).
\label{eq:ckm-m}
\end{equation}
The CKM matrix elements represent the coupling strength through $W^\pm$ boson
weak interaction between up-type and down-type quarks. The diagonal elements,
which describe transitions within the same generation, are measured to be of
$\mathcal{O}(1)$, whereas the off-diagonal elements, which describe transitions
among different generations, are of $\mathcal{O}(10^{-2}\textrm{-}10^{-3})$.
In the case of charm physics, the $c$ quark is more likely to decay to the $s$
quark, called Cabibbo-favored, than to the $d$ quark, called
Cabibbo-suppressed.

\begin{figure*}[hbp]
  \centering
  \subfigure[]{\includegraphics[]{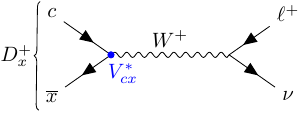}}\hskip -1pt
  \subfigure[]{\includegraphics[]{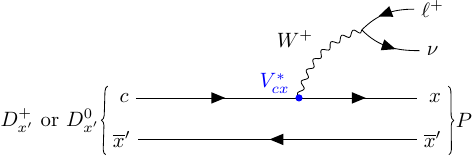}}
  \caption{Diagram of the purely leptonic (a) and semi-leptonic (b) decay of a
    $D$ or $D_s$ meson. Quark $x$ is a $d$ or an $s$ quark, and $x'$ is either
    $u$, $d$ or $s$. The corresponding element of the CKM matrix, $V_{cx}$,
    appears in the vertex.}
  \label{fig:diagram}
\end{figure*}

The unitarity of $V_{\rm CKM}$ is a fundamental requirement of SM. Any
deviation will reveal the sign of new physics. Consequently, a principal goal
of flavor physics is to search evidence of the deviation with breakthrough in
precision of the CKM matrix element measurements. The precision at which the
elements in the first row, $|V_{ud}|$ and $|V_{us}|$, are known has reached
$1\times10^{-4}$-$1\times10^{-3}$, but that of the second row, $|V_{cs}|$ and
$|V_{cd}|$, is only $\mathcal{O}(10^{-2})$ at present~\cite{PDG}. Improving
the precision of the determination of the elements in the second row,
therefore, will effectively enhance the sensitivity of testing the unitarity of
CKM matrix. These elements in the charm sector, $|V_{cs}|$ and $|V_{cd}|$, can
be accessed by the purely leptonic and semi-leptonic charmed-meson decays. In a
purely leptonic decay of a $D_{(s)}^+$ meson, which is shown in the Feynman
diagram in Fig.~\ref{fig:diagram}(a), the $c$ quark and the $d$~($s$)
quark annihilate, followed by a weak current connecting to the system of the
lepton $\ell^+$ and the corresponding flavored neutrino
$\nu_\ell$~($\ell=e,\mu,\tau$). The Feynman diagram in
Fig.~\ref{fig:diagram}(b) depicts the semi-leptonic decay, where the
initial-state charmed hadron evolves to the final-state hadrons along with a
weak current emitted externally. Their decay rates are propotional to the
product of the CKM matrix element $|V_{cd(s)}|$ and the decay constant or form
factors (see Eq.~\eqref{eq01} in section~\ref{sec:decay_constants}, and
Eq.~\eqref{eq:semi} in section~\ref{sec:form_factors}). Measurements of decay
rates for the purely leptonic and semi-leptonic decays can be readily used to
study weak-interaction physics.

The CKM matrix can be parameterized by three mixing angles and one CP-violating
complex phase~\cite{Chau:1984fp}. The SM coupling strengths for the $u-s$ and
$c-d$ transitions are both equal to $G_{F}|\sin\theta_c|$ with a small, well
understood, $\mathcal{O}(10^{-4})$ correction. Here $G_{F}$ is the Fermi
constant, and $\theta_c$, one of the three mixing angles, is called the Cabibbo
angle. Any significant difference in $|\sin\theta_c|$ extracted from different
quark transitions would be an unambiguous sign of new physics. The measured
accuracy of about 0.2\% level from nuclear $\beta$ and kaon decays for
$|V_{ud}|$ and $|V_{us}|$, is more than an order of magnitude better than those
from $D_s$ and $D$ decays for $|V_{cs}|$ and $|V_{cd}|$, which is about $ 3\%$,
based on statistics-limited BESIII measurements of $D_s^+\to \mu^+\nu$,
$D^+\to \mu^+\nu$ and $D^+\to K^{-}(\pi^{-})\ell^+\nu_\ell$ decays. The clean
environments for $D^{+(0)}$ and $D_s^+$ mesons produced by
$e^+e^- \to \psi(3770)\to D\bar{D}$ and $\psi(4160)\to D_s^{*\pm}D_s^\mp$,
respectively, are especially well suited for low-systematic-uncertainty
$c$-quark transition measurements. Year-long runs of future super tau-charm
factory~\cite{SCTF_LoI,STCF_LoI} at $\sqrt{s}=$3.773~GeV and 4.160~GeV, which
correspond to the rest masses of $\psi(3770)$ and $\psi(4160)$ respectively, 
would reduce the statistical uncertainties on $c$-quark-related determinations
of $|\sin\theta_c|$ to the level of about  0.2\%, and match those from $\beta$
and Kaon decays for the measurements of $|V_{ud}|$ and $|V_{us}|$.

\section{Experimental overview}
Charmed hadrons can be produced at a range of different
accelerators~\cite{Gersabeck:2015wna}. The production mechanisms are different,
and the production cross-sections are largely varied. 

The production cross-sections at hadron colliders are significantly higher than
those at $e^+e^-$ colliders. For example, the cross-section for producing
$c\bar{c}$ pairs in proton-proton collisions at the LHC with $\sqrt{s}=7$~TeV
is more than a million times higher compared to operating an $e^+e^-$ collider
at the $\psi(3770)$ and $\Upsilon(4S)$  resonances of the charm and $B$
factories~\cite{BaBarBelle_Xsec,BES_Xsec,LHCb_Xsec}. For high energy hadron
collisions, the fractional energy of the two colliding partons is likely to be
different. This difference leads to an energy boost of the charmed hadrons of
the collision system. Therefore, LHCb is ideally suited for performing
decay-time dependent studies of charm decays. However, it is more difficult to
study the leptonic and semileptonic decays of charmed hadrons at hadron
colliders, as there are fewer kinematic constraints.

At $e^+e^-$ colliders, there are more kinematic constraints available since the
collision energy of the initial electron and positron is well known. The
trigger efficiency of hadronic events at $e^+e^-$ colliders can reach almost
$100\%$, and with a relatively clean environment compared to hadron colliders.
Two different running conditions can be operated for studying the charm
physics. Running at a higher $\sqrt{s}$ to produce $\Upsilon(4S)$ on resonance
is used by the $B$ factories: BaBar and Belle$/$Belle II experiments, which are
located at the PEP-II and KEKB$/$Super KEKB colliders, respectively. Both
PEP-II and KEKB$/$Super KEKB are asymmetric colliders, which means that the
energies of the $e^+$ and $e^-$ beams differ. Hence, they have a collision
system that is boosted with respect to the laboratory frame. This allows
decay-time dependent studies for charmed hadrons. The alternative is tuning
$\sqrt{s}$ to slightly more than the production threshold of
charmed~(anti-charmed) hadron pairs. Since there is no extra energy for
producing pions, the decay  goes almost exclusively through charmed hadron
pairs. This production mode has been used for the CLEO-c experiment at the
CESR-c collider as well as for BESIII at BEPCII, which is ideally suited for
the studies of leptonic and semileptonic decays of charmed hadrons. However, it
is worth mentioning that the decay-time dependent studies of charmed hadrons
are not possible for these experiments since the collisions happen at rest.

BESIII has collected data samples with integrated luminosity of 8.0~fb$^{-1}$ 
at center-of-mass energy~($\sqrt{s}$) 3.773~GeV, 7.33~fb$^{-1}$ at
$\sqrt{s}=$4.128-4.226~GeV, and 4.5~fb$^{-1}$ between $\sqrt{s}=$4.600 and
4.699~GeV; these $\sqrt{s}$ values are optimal for the accumulation of
$D\bar{D}$, $D^{*+}_{s}D_{s}^{-}$, and $\Lambda^{+}_{c}\bar{\Lambda}^{-}_c$
events near thresholds, respectively. BESIII will ultimately collect data
samples with integrated luminosities of 20~fb$^{-1}$ at $\sqrt{s}=$3.773~GeV
and 6~fb$^{-1}$ around 4.18~GeV in the near future. The kinematics at
threshold allows the undetected neutrinos to be reconstructed according to the
four-momentum conservation. Charmed hadron (semi-)leptonic measurements with
these data samples provide rigorous tests of QCD, CKM unitarity, and lepton
flavor universality~(LFU) that complement similar studies of beauty hadrons.
For charmed baryons, the measurement of the branching fraction of 
$\Lambda^+_c \to \Lambda \ell^+ \nu_\ell$ was first performed by the ARGUS 
collaboration~\cite{ARGUS:1991bvx} and subsequently by the CLEO 
collaboration~\cite{CLEO:1994lyw}. Recently, the BESIII Collaboration 
measured the absolute branching fraction of 
$\Lambda^+_c \to \Lambda \ell^+ \nu_\ell$~\cite{2022_Lamdbac_Lenu, Ablikim:2016vqd}.
In addition to precision branching fractions of the $\Lambda_c$, if the BEPCII
c.m.~energy were upgraded to reach above 4.95 GeV, slightly beyond the current
plan of 4.9~GeV, pairs of $\Sigma_c$ and $\Xi_c$ baryons could also be produced 
and studied.

BEPCII delivered its first physics data set in 2009 on the $\psi(2S)$ peak.
Since then, BESIII has collected more than 40~fb$^{-1}$ of integrated 
luminosity at different $\sqrt{s}$ from 2.0 to 4.95~GeV. In order to extend the
physics potential of the BESIII experiment, two upgrade plans for BEPCII were
proposed and approved in 2020. The first upgrade will increase the maximum beam
energy to 2.8~GeV (corresponding to $\sqrt{s}=$5.6~GeV), which will expand the
energy reach of the collider into new territory. The second upgrade will
increase the peak luminosity by a factor of 3 for $\sqrt{s}=4.0$-$5.6$~GeV. It
will take about 2.5 years to prepare the upgraded components and half a year
for installation and commissioning, which will start in June 2024 and finish in
December 2024. With these upgrades, BESIII will enhance its capabilities to
explore physics of the exotic charmonium-like states~(XYZ) and will have the
unique ability to perform precision measurements of the production and decays
of charmed mesons and baryons at different thresholds. Recently, two $e^+e^-$
collider experiments are proposed in Russia (SCTF) and China (STCF), which will
act as super Charm factories~\cite{SCTF_LoI,STCF_LoI} with much improved
luminosity and broader energy range for BESIII experiment.

\section{Advantage of threshold production of charm hadron pairs }
\label{sec:DT}
The neutrinos produced in (semi-)leptonic processes cannot be directly detected.
However, the kinematics of threshold production of the charmed hadron pairs allow a
tag technique
to be employed, which provides a unique, low-background environment to measure 
the absolute branching fractions for charmed hadrons decaying to leptonic and
semi-leptonic final states although it contains one undetectable particle.

The $\psi(3770)$ resonance is predominantly decays to final states with a 
$D\bar{D}$ meson pair ($D\bar{D}$ represents $D^+D^-$ or $D^0\bar{D}^0$). The
leftover phase-space is only 30-40~MeV/$c^2$, which is 
not even enough for the lightest hadron, $\pi$. Hence, when a $D$ 
(or $\bar{D}$) meson is reconstructed via one of its hadronic decay modes, the 
other $\bar{D}$ (or $D$) must be the parent of all the remaining particles in 
the event, even if some of them escape from the detector. In the case of (semi-)leptonic 
decays, the mass of the missing neutrino can be inferred from the four-momentum 
conservation. This tagging feature is a robust tool for charmed particle decay 
measurements in near threshold experiments.

There are two types of samples used in the tag technique: single tag~(ST) and
double tag~(DT). In the ST sample, a charmed hadron (such as a $D$ meson)  is
reconstructed through
a particular hadronic decay without any requirement on the remaining reconstructed particle candidates. 
To select or extract the signal yields of the ST sample,
beam constrained mass $M_{\rm BC}$ and energy difference $\Delta E$ are the 
variables usually used:
\begin{equation}
    M_{\rm BC}=\sqrt{E_{\rm beam}^2-|\vec{p}_{D}|^2}\,, \,\,\,\,
    \Delta E=E_{D}-E_{\rm beam}\,,
\end{equation}
where $|\vec{p}_{D}|$ and $E_{D}$ are the total reconstructed momentum
and energy of the $D$ candidate in the center-of-mass frame
of the $\psi(3770)$ resonance, respectively, and $E_{\rm beam}$ is the calibrated
beam energy. The $D$ signals will be consistent with the
nominal $D$ mass in $M_{\rm BC}$ and with $\Delta E=0$.

In the DT sample, events can be fully
reconstructed. The $D$ hadron of the pair, designated as the
``tag'' side, is usually reconstructed through 
well-measured hadronic decay channels,
while the anti-$D$-hadron of the pair, designated as the ``signal'' side, can be
reconstructed through the (semi-)leptonic decay mode of interest.
The information of the undetectable particle, e.g.~neutrinos,
are accessed via the missing four-momentum.  

The yield of a ST sample of a tag mode is given by
\begin{equation}\label{eq:ST_yield}
  N^{\text{ST}}_{\text{tag}} = 2N_{D\bar{D}}{\cal B}_{\text{tag}}\varepsilon_{\text{tag}}\,,
\end{equation}
and the yield of a DT sample is given by
\begin{equation}\label{eq:DT_yield}
  N^{\text{DT}}_{\text{tag,sig}} = 2N_{D\bar{D}}{\cal B}_{\text{tag}}{\cal B}_{\text{sig}}\varepsilon_{\text{tag,sig}}\,,
\end{equation}
where $N_{D\bar{D}}$ is the total number of produced $D\bar{D}$ pairs,
${\cal B}_{\text{tag(sig)}}$ is the branching fraction of the tag (signal) side,
and the $\varepsilon$ are the corresponding efficiencies. The branching fraction of the signal decay under study can be determined
by isolating ${\cal B}_{\text{sig}}$ such that
\begin{equation}\label{eq:DTag_BF}
  {\cal B}_{\text{sig}} = \frac{N^{\text{DT}}_{\text{tag,sig}}}{N^{\text{ST}}_{\text{tag}}}\frac{\varepsilon_{\text{tag}}}{\varepsilon_{\text{tag,sig}}}\,.
\end{equation}

Another advantage of the tag technique is that the efficiency for
reconstructing the tag-side should almost cancel and any residual effects 
caused by the tag-side are expected to be negligible since the efficiency 
approximately factorizes: 
$\varepsilon_{\text{tag,sig}}\,\approx\,\varepsilon_{\text{tag}}\varepsilon_{\text{sig}}$.
The residual effects might occur when the factorization is violated. In other
words, the DT efficiency is not completely the product of both sides' ST 
efficiencies, since ST candidates are measured with the other $D$ hadron of the
pair decaying generically, while the tag side of DT candidates are measured
versus a specific decay mode. The numbers and types of the decay products of
the other $D$ hadron matters due to overlapping on the detector. This effect,
however, is believed to be small due to the fine-grained BESIII
detector.\footnote{The violations of factorization are not negligible, but the
effect is taken into account in the Monte-Carlo simulation. In practice, the
systematic uncertainty caused by tag sides is about 0.1\%.}

The $D_{s}^{+}$ samples are mainly obtained from the dataset at 
$\sqrt{s}=$4.128-4.226~GeV through the process 
$e^{+}e^{-}\to D_{s}^{*\pm}D_{s}^{\mp}\to \gamma/\pi^0 D_{s}^{+}D_{s}^{-}$.
\footnote{
The threshold of $D_{s}^{\pm}D_{s}^{\mp}$ pairs is at
about $\sqrt{s}=$4.009~GeV, while the cross section is only about one third of
that for $e^{+}e^{-}\to D_{s}^{*\pm}D_{s}^{\mp}$ at $\sqrt{s}=$4.180~GeV.
}
Unlike $D\bar{D}$ pairs, $D_{s}^{+}D_{s}^{-}$ pairs don't evenly share the 
four-momentum of the $e^{+}e^{-}$ system. Accordingly, the recoil mass of the
tag $D_{s}$ is a popular variable to select signal events while the tag
technique is still a robust tool.

\section{Lattice QCD overview}
Given experimental measurements of the branching fractions of charmed hadron decays, 
 combined with sufficiently precise theoretical calculations of the hadronic matrix
elements, together they enable the determination of the CKM matrix elements $|V_{cd}|$ 
and $|V_{cs}|$ and a precise test of the unitarity of the second row of the
CKM matrix. Significant progress has been made in charm physics on the
lattice in recent years, largely due to the availability of gauge configurations produced
using highly-improved lattice-fermion actions that enable treating the $c$ quark with the
same action as for the $u$, $d$, and $s$ quarks. Different LQCD collaborations 
use different discretizations of the gluon and fermion actions, which is important for
cross-checks to uncover potential systematic effects. The analysis techniques used
by different groups may also differ, and the results should only agree after the 
physical/continuum limit has been taken.  Indeed, the consistency of values
obtained using different formulations adds significantly to our confidence in the results.

In this review we use lattice results reported by the Flavour Lattice Averaging Group (FLAG)
in their 2021 review~(published in 2022)~\cite{FlavourLatticeAveragingGroupFLAG:2021npn}.
FLAG has a set of guidelines to combine current, published LQCD results, including only
lattice calculations for which all systematic effects have
been taken into account, into a
world average. In some cases only a single result enters the average,
but more often it is a combination of results from different collaborations. The
lattice results are grouped and listed according to the number of sea-quark flavors
used in the calculation. Here we take the $n_f=2+1+1$ results, that include $u/d$, 
$s$ and $c$ quarks in the sea, if these results are available --- otherwise we
take the $n_f=2+1$ results. The 2 here means that $u$ and $d$ quarks are treated as
degenerate on the lattice, which helps in keeping the computational cost of the calculations
reasonable. For charm decays, this is estimated to be a small effect and can be taken
into account. Calculations that include $n_f=1+1+1+1$ flavors of sea quarks,
treating $u$ and $d$ quarks as non-degenerate, are now emerging for quantities where
the effect is expected to be most important.

\section{Purely leptonic and semi-leptonic decays and lepton flavor universality} 
One of the most charming feature of charmed mesons is that their $\sim 2$~GeV 
masses locate them in the region where nonperturbative QCD is activated. This 
fact is an obstacle to study their hadronic transitions, but their purely
leptonic and semi-leptonic decays play an important medium to facilitate such
investigation. Purely leptonic and semi-leptonic decays of $D_{(s)}^+$ provide
the cleanest and best access to understand the mechanism of the $c$ quark to
the $d$($s$) quark transition (Fig.~\ref{fig:diagram}). In both decays, because
strong interactions affect only the hadronic system, the weak and strong
interactions can be well separated allowing the most precise determinations of
corresponding decay constants (Sec.~\ref{sec:decay_constants}), form factors
(Sec.~\ref{sec:form_factors}), and CKM matrix elements (Secs.~\ref{sec:CKM}
and~\ref{sec:discussionVcdVcs}).

\subsection{Leptonic decays and decay constants for $D^+$ and $D_s^+$} 
\label{sec:decay_constants}
The leptonic decay of the charmed meson $D_{(s)}^{+}$ could happen by the
annihilation of the $c$ and $\bar{d}(\bar{s})$ quarks into $\ell^+\nu_\ell$
mediated by a virtual $W^+$ boson. The partial decay width of the
$D_{(s)}^+ \to \ell^+\nu_\ell$ to the lowest order within SM, due to the
isolation of the weak and strong effects, is proportional to the product of
the decay constant $f_{D_{(s)}^+}$ (parametrizing strong-interaction effects
between the two initial-state quarks) and the CKM matrix element
$|V_{cd(s)}|$~(parametrizing the $c\to s(d)$ falvor-changing interaction),
given in a simple form:~\cite{PhysRevD.38.214}
\begin{equation}
\Gamma(D_{(s)}^+ \to \ell^+\nu_\ell)=
     \frac{G^2_F f^2_{D_{(s)}^+}} {8\pi}
       |V_{cd(s)} |^2
      m^2_\ell M_{D_{(s)}^+}
    \left (1- \frac{m^2_\ell}{M^2_{D_{(s)}^+}}\right )^2,
\label{eq01}
\end{equation}
\noindent
where $m_\ell$ and $M_{D_{(s)}^+}$ stand for the $\ell$ lepton and the
$D_{(s)}^+$ meson invariant mass, respectivley. The partial decay width is
calculated by the branching fraction (Eq.~\ref{eq:DTag_BF}) divided by the $D_{(s)}^+$ life time obtained through global fit~\cite{PDG}.
The decay constants $f_{D_{(s)}^+}$ can be calculated using
LQCD, while the CKM matrix elements can only be determined by experimental measurements.
Measuring the partial decay width (or branching fraction) of $D_{(s)}^+ \to \ell^+\nu_\ell$, one
can either determine the $f_{D_{(s)}^+}$ by taking the
$|V_{cd(s)}|$ from the SM global fit~\cite{CKMFitter}, or extract
the $|V_{cd(s)}|$ with the help of the $f_{D_{(s)}^+}$ predicted by LQCD.
The precision of the $f_{D_{(s)}^+}$ or $|V_{cd(s)}|$ determination is directly
propagated from the precision of the $D_{(s)}^+ \to \ell^+\nu_\ell$
measurement, since
the uncertainties associated with the higher-order correction of the SM model
and the precisely measured $G_F$, $m_\ell$ and $M_{D_{(s)}^+}$ are negligible with current statistics. 

In LQCD calculations, the decay constants $f_{D_{(s)}}$ are extracted 
from Euclidean matrix elements of the axial current
\begin{equation}
\langle 0|A^\mu_{cx}|D_x(p)\rangle = if_{D_x}p^\mu\,,
\end{equation}
with $x = d$ or $s$ quark, $A^\mu_{cx}=\bar{c}\gamma_\mu \gamma_5 x$ and $p$ momentum
of the $D_x$ meson. Results for $n_f = 2+1+1$ dynamical flavors from the FLAG review~\cite{FlavourLatticeAveragingGroupFLAG:2021npn}
are summarized in Table~\ref{tab:pure}  
alongside experimental results. In the case of LQCD, we list isospin-averaged 
quantities, although, in a few cases, results for $f_{D_+}$ have been published  \cite{FermilabLattice:2011njy,FermilabLattice:2014tsy,Bazavov:2017lyh}. In one of the lattice calculations, the difference between $f_D$ and $f_{D^+}$ has been estimated to be $0.58\pm 0.07$ MeV~\cite{Bazavov:2017lyh}. 
The results for the decay constants from LQCD are very
precise, the total uncertainties (including both statistical and systematic 
uncertainties) for the FLAG averages being 0.33\% and 0.2\% for $f_D$ and $f_{D_s}$,
respectively.

\begin{table}[htp]
 \renewcommand\arraystretch{1.25}
\centering
\caption{\label{tab:pure}
Measurements of $D^+$ and $D^+_s$ purely leptonic decays with threshold data at 
BESIII,  and comparisons between experimental results and theoretical 
expectation or SM-global fit results. (Here, ``-" indicates not available.)
For theory predictions of the decay constants, we take the
Flavour Lattice Averaging Group (FLAG) averages~\cite{FlavourLatticeAveragingGroupFLAG:2021npn} of calculations with $n_f=2+1+1$ sea quarks.
The original calculations included in the average are cited next to the results. Note
that these predictions, which are marked with an asterisk in the table, are for the 
isospin-averaged quantities $f_D$ and $f_{D_s}$. In a few cases, results for $f_{D^+}$ have been published~\cite{FermilabLattice:2011njy,FermilabLattice:2014tsy,Bazavov:2017lyh}, and the 
difference between $f_D$ and $f_{D^+}$ has been estimated to be 
$0.58\pm 0.07$ MeV \cite{Bazavov:2017lyh}. We add this to the isospin-averaged quantity and combine 
the lattice estimate for $f_D$ with the experimental result for $f_{D^+}|V_{cd}|$ to get
an estimate for the CKM element given in the "Measurement" column. For $f_{D_s}$
we assume the difference between $f_{D_s}$ and $f_{D^+_s}$ is the same $0.58$ MeV,
but double the uncertainty to be conservative in estimating the uncertainty. The values
given for $V_{cs}$ in the "Measurement" column are then calculated from the
experimental result for $f_{D^+_s}|V_{cs}|$ and the LQCD result for the decay constant.
For comparison, we give the values of the CKM elements extracted by the CKMfitter group~\cite{CKMFitter} 
in the "Prediction/Fit" column.}
\begin{tabular}{lccc}
\hline\hline
Observable & Measurement  & Prediction/Fit \\ \hline
$\mathcal B(D^+\to \mu^+\nu_\mu)$& $(3.71 \pm 0.19_\mathrm{stat}\pm 0.06_\mathrm{syst}) \times 10^{-4}$~\cite{bes3_muv} &- \\
$f_{D^{+}}|V_{cd}| $ & $(45.75 \pm 1.20_\mathrm{stat}\pm 0.39_\mathrm{syst})$ MeV&-  \\
$f_{D^{+}} $  &   $(203.8 \pm 5.2_\mathrm{stat} \pm 1.8_\mathrm{syst})$ MeV & $ (212.0 \pm 0.7)$ MeV$^\ast$~\cite{Bazavov:2017lyh, Carrasco:2014poa}\\
$|V_{cd}| $ & $0.2152 \pm 0.0060$ & $0.22487^{+0.00024}_{-0.00021}$\\ \hline
\hline
$\mathcal B(D^+ \to \tau^+(\pi^+\bar{\nu}_\tau)\nu_\tau) $ &  $(1.20 \pm 0.24_\mathrm{stat}\pm 0.12_\mathrm{syst}) \times 10^{-3}$~\cite{Ablikim:2019rpl} &- \\  \hline
$\Gamma(D^+ \to \tau^+ \nu_\tau)/  \Gamma(D^+ \to \mu^+ \nu_\mu)$ & $3.21 \pm 0.74$~\cite{Ablikim:2019rpl} &  2.67 \\ \hline
\hline
$\mathcal B(D^+_s \to \mu^+\nu_\mu)$& $(5.35 \pm 0.13_\mathrm{stat}\pm 0.16_\mathrm{syst})\times 10^{-3}$~\cite{bes3_Ds_tauv2} &-  \\
$f_{D^+_s}|V_{cs}| $ & $(243.1 \pm 3.0_\mathrm{stat}\pm 3.7_\mathrm{syst})$ MeV & -  \\
$f_{D^+_s} $  &   $( 249.8 \pm 3.0_\mathrm{stat} \pm 3.8_\mathrm{syst})$ MeV & $ (249.9 \pm 0.5)$ MeV$^\ast$~\cite{Bazavov:2017lyh, Carrasco:2014poa} \\
$|V_{cs}| $ &  $0.971 \pm 0.019$ & $0.973521^{+0.000057}_{-0.000062}$ \\ 
\hline
$\mathcal B(D_s^+\to\tau^{+}(\pi^+\bar{\nu}_\tau)\nu_\tau)$& $(5.22 \pm 0.25_\mathrm{stat}\pm 0.17_\mathrm{syst})\%$~\cite{bes3_Ds_tauv2} &-  \\
$f_{D^+_s}|V_{cs}| $ & $(243.0 \pm 5.8_\mathrm{stat}\pm 4.0_\mathrm{syst})$ MeV & -  \\
$f_{D^+_s} $  &   $(  249.7\pm 6.0_\mathrm{stat} \pm 4.2_\mathrm{syst})$ MeV & $ (249.9 \pm 0.5)$ MeV$^\ast$~\cite{Bazavov:2017lyh, Carrasco:2014poa} \\
$|V_{cs}| $ &  $0.970 \pm 0.028$ & $0.973521^{+0.000057}_{-0.000062}$ \\ 
\hline
$\mathcal B(D_s^+\to\tau^{+}(\rho^+\bar{\nu}_\tau)\nu_\tau)$& $(5.29 \pm 0.25_\mathrm{stat}\pm 0.20_\mathrm{syst})\%$~\cite{bes3_Ds_tauv1} &-  \\
$f_{D^+_s}|V_{cs}| $ & $(244.8 \pm 5.8_\mathrm{stat}\pm 4.8_\mathrm{syst})$ MeV & -  \\
$f_{D^+_s} $  &   $(  251.6\pm 5.9_\mathrm{stat} \pm 4.9_\mathrm{syst})$ MeV & $ (249.9 \pm 0.5)$ MeV$^\ast$~\cite{Bazavov:2017lyh, Carrasco:2014poa} \\
$|V_{cs}| $ &  $0.977 \pm 0.030$ & $0.973521^{+0.000057}_{-0.000062}$ \\ 
\hline
$\mathcal B(D_s^+\to\tau^{+}(e^+\nu_e\bar{\nu}_\tau)\nu_\tau)$& $(5.27 \pm 0.10_\mathrm{stat}\pm 0.12_\mathrm{syst})\%$~\cite{bes3_Ds_tauv3} &-  \\
$f_{D^+_s}|V_{cs}| $ & $(244.4 \pm 2.3_\mathrm{stat}\pm 2.9_\mathrm{syst})$ MeV & -  \\
$f_{D^+_s} $  &   $(  251.1\pm 2.4_\mathrm{stat} \pm 3.0_\mathrm{syst})$ MeV & $ (249.9 \pm 0.5)$ MeV$^\ast$~\cite{Bazavov:2017lyh, Carrasco:2014poa} \\
$|V_{cs}| $ & $0.976 \pm 0.015$ & $0.973521^{+0.000057}_{-0.000062}$ \\ 
\hline
$\Gamma(D^+_s \to \tau^+ \nu_\tau)/  \Gamma(D^+_s \to \mu^+ \nu_\mu)$ & $9.80\pm 0.34$~\cite{PDG} & 9.75\\ \hline
\hline
\end{tabular}
\end{table}

In recent years, the BESIII collaboration has reported the most precise
experimental studies of $D_{(s)}^+ \to \ell^+\nu_\ell$ by using $e^+e^-$
annihilation data corresponding to a total integrated luminosity of 2.93, 0.48,
and 6.32~fb$^{-1}$ collected at
$\sqrt{s}=$3.773~GeV for $D^{0}\bar{D}^{0}/D^{+}D^{-}$, 4.009~GeV for $D_{s}^{+}D_{s}^{-}$,
and 4.178-4.226~GeV for $D_{s}^{*\pm}D_{s}^{\mp}$,
respectively~\cite{bes3_muv,Ablikim:2019rpl,bes3_Ds_muv,bes3_Ds_tauv1,bes3_Ds_tauv2,bes3_Ds_tauv3}.

With the $3.773$~GeV data sample, BESIII measured the branching fraction of
$D^+\to\mu^+\nu_\mu$~\cite{bes3_muv}, and either 
used lattice results on the decay constant $f_{D^+}$ to extract the CKM matrix element $|V_{cd}|$ or used global fit of $|V_{cd}|$ to obtain $f_{D^+}$.
Furthermore, BESIII observed the
$D^+\to\tau^+(\pi^+\bar{\nu}_\tau)\nu_\tau$ decay for the first
time~\cite{Ablikim:2019rpl}. The corresponding results are summarized in
Table~\ref{tab:pure}. A total of about 1.7 million $D^{-}$ events are tagged through nine
$D^{-}$ hadronic decays (summing up to 30\% of all $D^{-}$ decays). Signal
candidates of $D^+\to\mu^+\nu_\mu$ or
$D^+\to\tau^+(\pi^+\bar{\nu}_\tau)\nu_\tau$ are then reconstructed recoiling
against the tagged $D^-$ mesons by requiring exactly one track identified as
$\mu^+$ or $\pi^+$.\footnote{The electron final state $D^+\to e^+\nu_e$ is suppressed due to helicity suppression.} 
The neutrino is undetectable, but the yields of the signal
decays can be extracted from the missing-mass-squared variable
${\rm MM}^2=(E_{\rm beam}-E_{\mu/\pi})^2-(-\vec{p}_{\rm tag}-\vec{p}_{\mu/\pi})^2$,
where $E_{\mu/\pi}$ and $\vec{p}_{\mu/\pi}$ ($\vec{p}_{\rm tag}$) are the energy and
three-momentum of the reconstructed $\mu^+$ or $\pi^+$ (tagged $D^-$ mesons). The $D^+\to\mu^+\nu_\mu$
signal events have only one neutrino missing and peak around ${\rm MM}^2=0$,
which represents the negligible invariant mass of neutrinos.
In the case of 
$D^+\to\tau^+(\pi^+\bar{\nu}_\tau)\nu_\tau$, the two missing neutrinos push the 
signal distribution toward ${\rm MM}^2>0$ and lead to a more complex 
background than that in the $D^+\to\mu^+\nu_\mu$ decay. 

\begin{figure}[hbpt]
  \centering
  \subfigure[]{\includegraphics[width=3.0in]{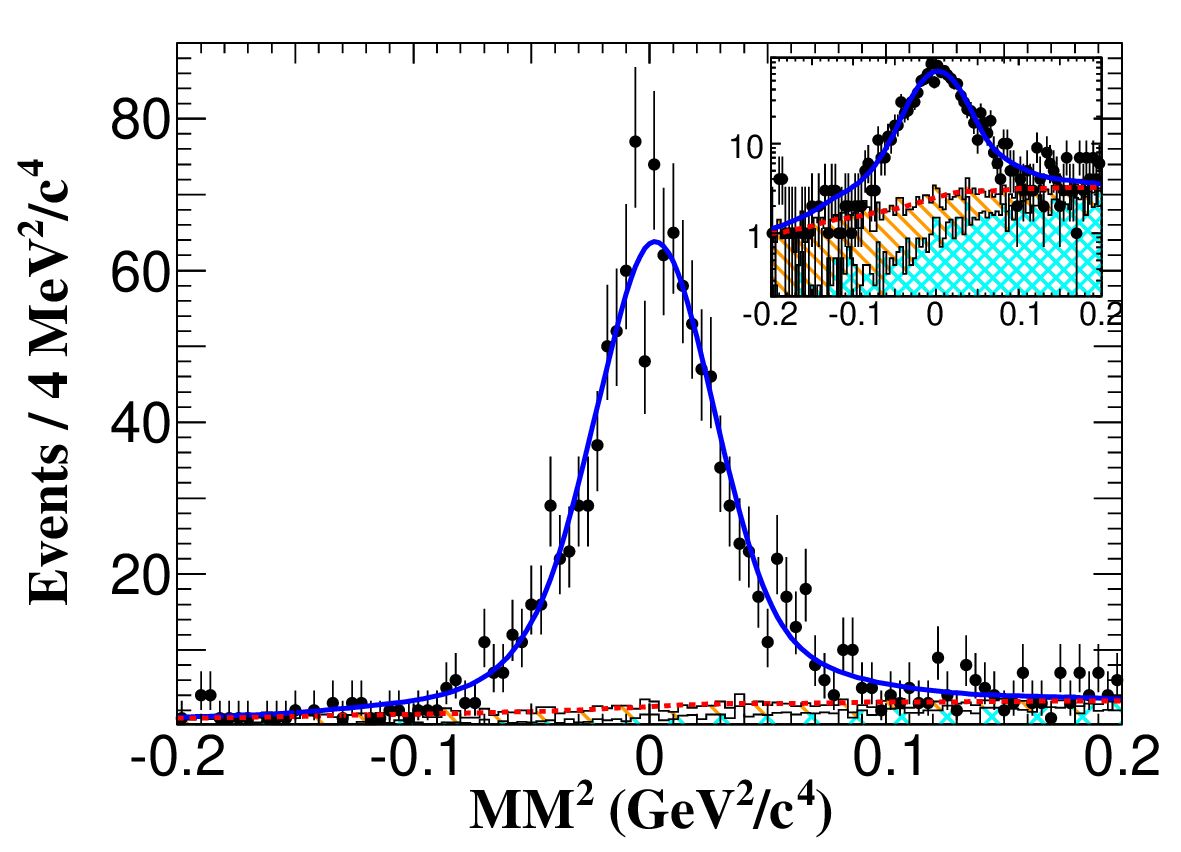}}
  \subfigure[]{\includegraphics[width=3.0in]{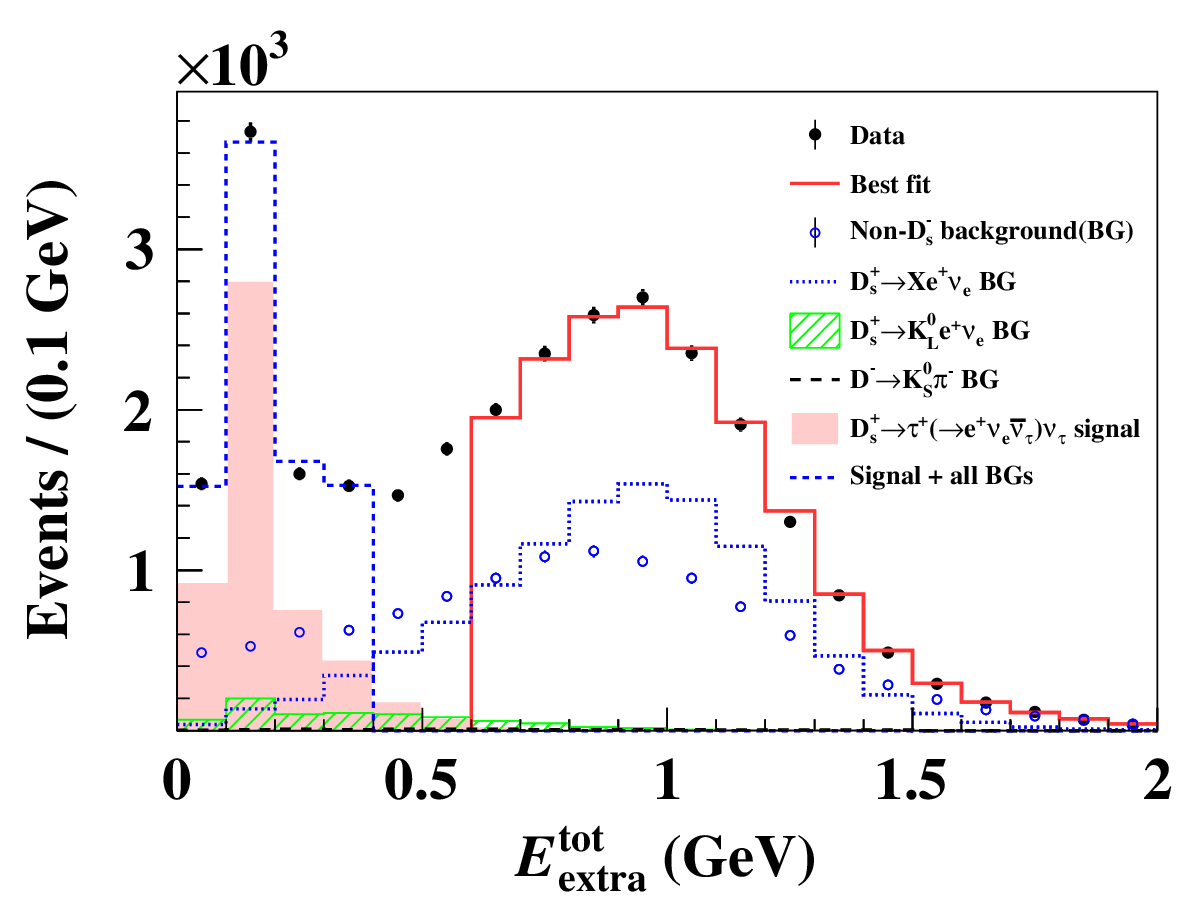}}
  \caption{(a) The squared missing mass~(${\rm MM}^2$) distribution of the selected
    $D^{+}_s \to \mu^{+} \nu_\mu$ candidates from Ref.~\cite{bes3_Ds_muv}.
    The error bars show the statistical uncertainty in experimental data. The
    inset shows the same distribution in logarithmic scale. Panel from
    Ref.~\cite{bes3_Ds_muv}. (b) The $E^{\rm tot}_{\rm extra}$ distribution of
    the selected $D^{+}_s \to \tau^{+}(e^+\nu_e\bar{\nu}_\tau) \nu_\tau$
    candidates from Ref.~\cite{bes3_Ds_tauv3}. The backgrounds, including
    non-$D_s^-$, $D_s^+ \to K_L^0 e^+ \nu_e$, and $D_s^+ \to X e^+ \nu_e$, are
    estimated by a fit to the region $E^{\rm tot}_{\rm extra}>0.6$~GeV and
    extrapolated into the signal region $E^{\rm tot}_{\rm extra}<0.4$~GeV using
    Monte-Carlo simulated shapes. Panel from Ref.~\cite{bes3_Ds_tauv3}.
  }
  \label{fig:dsenununu}
\end{figure}

With the $4.178\textrm{-}4.226$~GeV data samples, BESIII studied the 
$D_s^+\to\mu^{+}\nu_\mu$~\cite{bes3_Ds_muv,bes3_Ds_tauv2} and
$D_s^+\to\tau^{+}\nu_\tau$ decays with three $\tau^+$ decays: 
$\rho^+\bar{\nu}_\tau$~\cite{bes3_Ds_tauv1}, 
$\pi^+\bar{\nu}_\tau$~\cite{bes3_Ds_tauv2}, and 
$e^+\nu_e\bar{\nu}_\tau$~\cite{bes3_Ds_tauv3}. The measured branching 
fractions, decay constant $f_{D_s^+}$ and CKM matrix element $|V_{cs}|$ are 
listed in Table~\ref{tab:pure}. Using a similar method as for the $D^+$ purely 
leptonic decays, the 
$D_s^+\to\mu^{+}\nu_\mu$~(Fig.~\ref{fig:dsenununu}(a)),
$D_s^+\to\tau^{+}(\pi^+\bar{\nu}_\tau)\nu_\tau$, and
$D_s^+\to\tau^{+}(\rho^+\bar{\nu}_\tau)\nu_\tau$ signal yields are obtained 
from the ${\rm MM}^2$ distributions. However,
$D_s^+\to\tau^{+}(e^+\nu_e\bar{\nu}_\tau)\nu_\tau$ has three neutrinos 
missing, degrading ${\rm MM}^2$ into a non-ideal place to extract signal 
yields. Instead, the total energy of extra photons not used in event selection 
($E^{\rm tot}_{\rm extra}$) is employed to extract 
$D_s^+\to\tau^{+}(e^+\nu_e\bar{\nu}_\tau)\nu_\tau$ signal yields (Fig.~\ref{fig:dsenununu}(b)).
Different decay processes cause different numbers and energies of real/fake
photons inside the detector, and this decay causes a very distinct peak at low
$E^{\rm tot}_{\rm extra}$ where the backgrounds are small.

\begin{figure}[hbpt]
  \centering
  \subfigure[]{\includegraphics[width=3.0in]{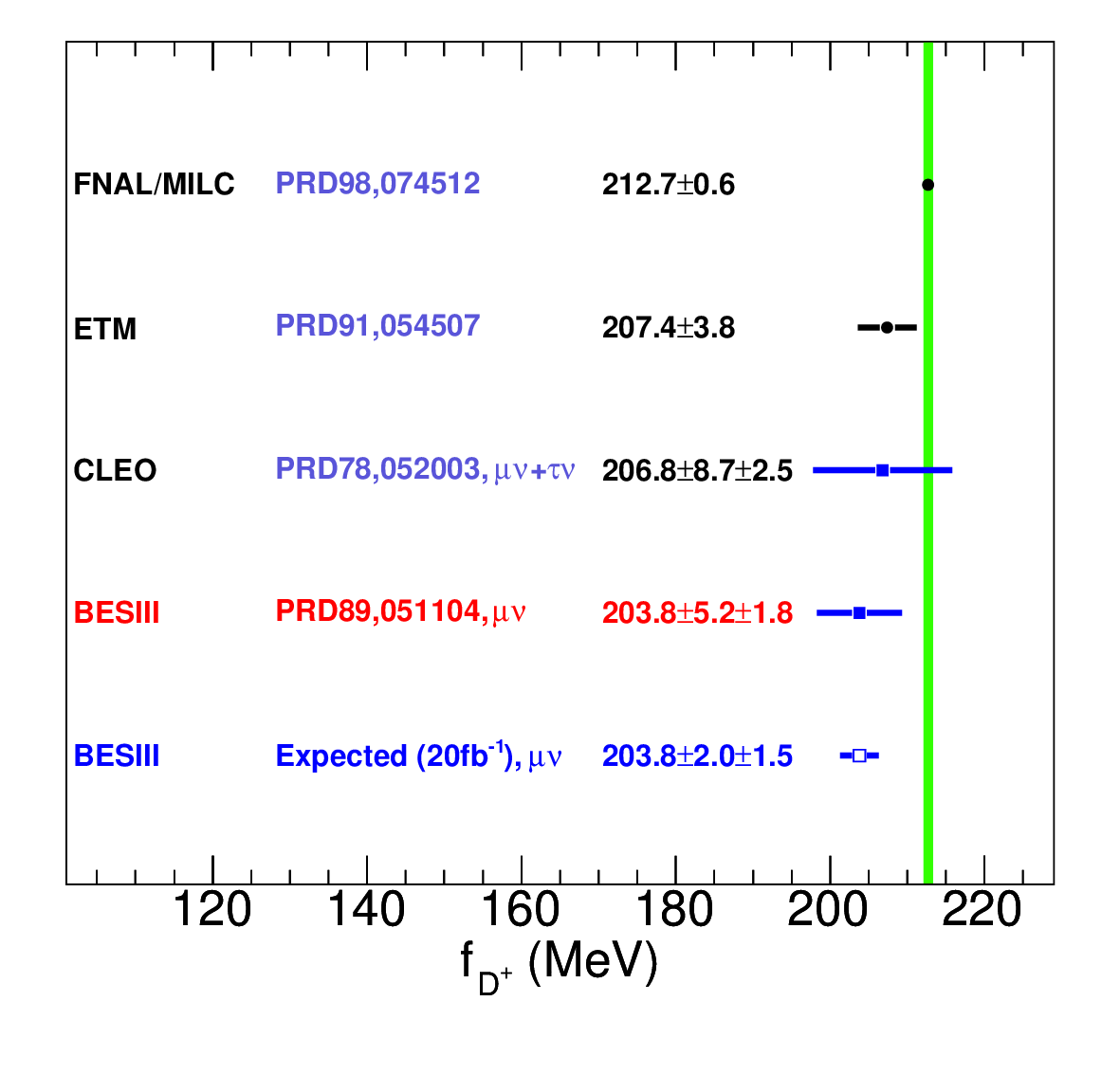}}\hskip -1pt
  \subfigure[]{\includegraphics[width=3.0in]{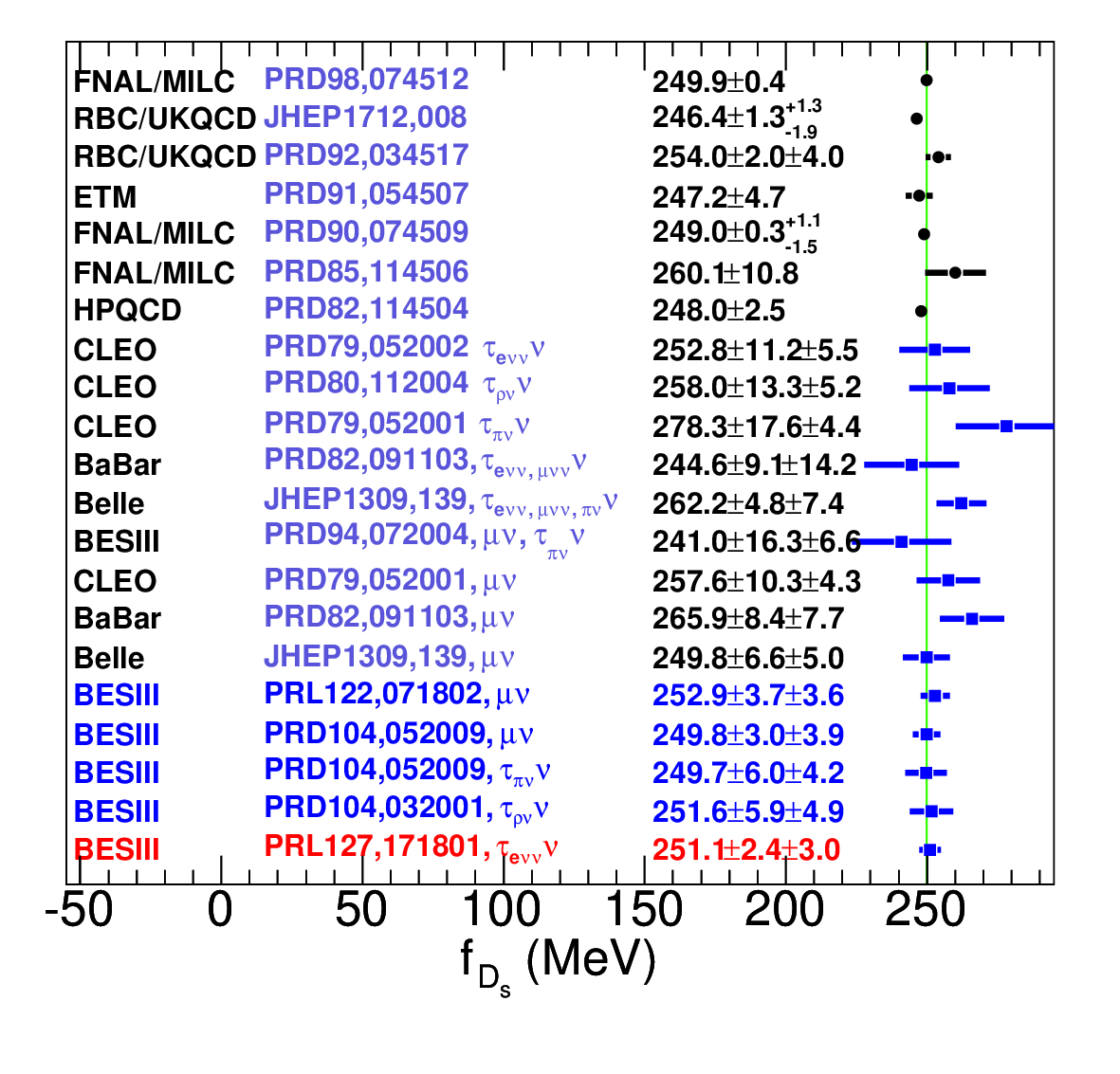}}
  \caption{Comparison of the results for (a) $f_{D^+}$ and (b)
    $f_{D_s^+}$  measured  by the BESIII, Belle, BaBar, and CLEO experiments. 
    The green bands present the LQCD uncertainties. Values marked in black
    circle denote LQCD calculations, values marked in blue square denotes
    experimental results, and the value marked in open square denotes the
    expected precision from BESIII with 20~fb$^{-1}$ data set at
    3.773~GeV~\cite{Ablikim:2019hff}. Note that the LQCD result for $f_D$ by
    ETM~\cite{Carrasco:2014poa} quoted in panel (a) is for the isospin-averaged
    quantity $f_D$, whereas the number quoted from
    FNAL/MILC~\cite{Bazavov:2017lyh} is for $f_{D^+}$. In panel (b) all lattice
    results are for the isospin-averaged quantity $f_{D_s}$. From these, ETM
    result~\cite{Carrasco:2014poa} and the top FNAL/MILC
    result~\cite{Bazavov:2017lyh} are $n_f=2+1+1$ calculations and included in
    the FLAG $n_f=2+1+1$
    average~\cite{FlavourLatticeAveragingGroupFLAG:2021npn}. All other LQCD
    results shown in panel (b) are $n_f=2+1$ calculations. Comparing these
    results shows that the effect of adding a charm quark to the sea is very
    small for $f_{D_s}$. Data from
    Refs.~\cite{bes3_muv, bes3_Ds_muv, bes3_Ds_tauv1, bes3_Ds_tauv2, bes3_Ds_tauv3, Bazavov:2017lyh, Boyle:2017jwu, Yang:2014sea, Carrasco:2014poa, FermilabLattice:2014tsy, FermilabLattice:2011njy, Davies:2010ip, CLEO:2009jky, CLEO:2009vke, CLEO:2009lvj, BaBar:2010ixw, Belle:2013isi, BESIII:2016cws,  CLEO:2008ffk}.
    Figure provided by Hai-Long Ma.
  } 
  \label{fig:decay_constant}
\end{figure}

The most precise measured $f_{D^+}|V_{cd}|$ value can be obtained via the
$D^+\to\mu^{+}\nu_\mu$ decay. The $D^+\to\tau^{+}(\pi^+\bar{\nu}_\tau)\nu_\tau$
contributes very little due to its large uncertainty. The most precise
measurement of $f_{D_s^+}|V_{cs}|$ comes from the $D_s^+$ purely leptonic
decays. In this case, $D_s^+\to\mu^{+}\nu_\mu$ and $D_s^+\to\tau^+\nu_\tau$
make comparable contributions (listed in Table~\ref{tab:pure}). Along with the
values of $|V_{cd}|$ and $|V_{cs}|$ from the SM-constrained fit~\cite{PDG},
$f_{D^+}$ and $f_{D_s^+}$ can be extracted. The comparisons of various
measurements are shown in Fig.~\ref{fig:decay_constant}. Furthermore, the Heavy
Flavor Averaging Group (HFLAV)~\cite{HFLAV:2022pwe}, having the measured
$f_{D_s^+}|V_{cs}|$ and $f_{D^+}|V_{cd}|$, gives a value
$f_{D_s}/f_D = 1.232 \pm 0.030$, which is roughly $1.8\sigma$ higher than the
lattice value from the FLAG world average value,
$1.1783 \pm 0.0016$~\cite{FlavourLatticeAveragingGroupFLAG:2021npn}.

\subsection{Semileptonic decays and transition form-factors}
\label{sec:form_factors}
In the SM, the theoretical description of semileptonic decays involve a 
leptonic current along with a hadronic current. The effects of these two
currents can be well-isolated~\cite{Ivanov:2019nqd} since the leptonic and
hadronic systems have no strong interactions between their final-states. The
hadronic current in the charm physics region is non-perturbative, and the
hadronic matrix element is thus well suited for a LQCD calculation.

The simplest case of the semi-leptonic $D_{(s)}^{0(+)}$ decays is
$D^{0/+} \to P \ell^+ \nu_\ell$ (where $P$ denotes a pseudoscalar meson).
Assuming that the pseudoscalar meson width is negligible,
the differential decay rate can be written as
\begin{equation}
\label{eq:fulldGammadq2}
\begin{split}
\frac{d\Gamma}{dq^2} =&\frac{G_F^2}{24\pi^3}|V_{cd(s)}|^2\frac{(q^2-m_\ell^2)^2\sqrt{E_P^2-M_P^2}}{q^4M^2_{D_{(s)}}}\\
&\times\left[\left(1+\frac{m_\ell^2}{2q^2}\right)M^2_{D_{(s)}}\left(E^2_P-M^2_P\right)|f_+(q^2)|^2+\frac{3m_\ell^2}{8q^2}\left(M^2_{D_{(s)}}-M_P^2\right)^2|f_+(q^2)|^2\right],
\end{split}
\end{equation}
where $E_P$ is the light-pseudoscalar meson energy in the rest frame of the decaying 
$D_{(s)}$, $M_P$ and $M_{D_{(s)}}$ are the meson masses, $m_\ell$ is the lepton
mass, and $q = (p_{D_{(s)}} - p_P)$ is the momentum of the outgoing lepton pair.
The vector and scalar form factors $f_+(q^2)$ and $f_0(q^2)$ parameterize the hadronic matrix element of the heavy-to-light quark flavour-changing vector current $V_\nu = \bar{d}\gamma_\mu c$ (or $V_\nu = \bar{s}\gamma_\mu c$),
\begin{equation}
\langle P|V_\mu|D_{(s)}\rangle = f_+(q^2)\left((p_{D_{(s)}})_\mu+(p_P)_\mu-\frac{M^2_{D_{(s)}}-M^2_P}{q^2}q_\mu\right)+f_0(q^2)\frac{M^2_{D_{(s)}}-M^2_P}{q^2}q_\mu
\end{equation}
and satisfy the kinematic constraint $f_+(0) = f_0(0)$. 

The contribution from
the scalar form factor to the decay width is proportonial to $m_\ell^2$, which
can usually be neglected. With this assumption, Eq.~\eqref{eq:fulldGammadq2} simplifies to
\begin{equation}
\label{eq:semi}
\frac{d\Gamma}{dq^2} =\frac{G_F^2}{24\pi^3}|V_{cs(d)}|^2
p_{P}^3 |f_{+}^{P}(q^2)|^2.
\end{equation}

From analyses of the dynamics in these semileptonic decays, one can obtain
the product of a form factor and a CKM matrix element: $|V_{cx}|f^P_+ (q^2)$.
One can either take $|V_{cd(s)}|$ from a global fit and extract the form
factor by choosing a parameterization (such as a simple pole and $z$-expansion),
or one can extract $|V_{cd(s)}|$ by using the form factor calculated in
LQCD. The shapes of the form factors from experiment and LQCD can
even be compared without knowledge of the CKM element, which simply appears
as a normalisation factor.

The CLEO experiment has made the first high precision measurements of
$D^{0/+} \to K/\pi e^+ \nu_e$ decay rates based on datasets accumulated at the
$\psi(3770)$ resonance peak with two different methods. The core concept
of the double tag method is to take the advantage of the pair production of
$D\bar{D}$ in the $\psi(3770)$ decays with the four-momentum conservation rule.
(Details are introduced in Section~\ref{sec:DT}.) 
On the tag side, one of the $D$ mesons is fully reconstructed 
via a hadronic decay mode, and on the signal side, the semi-leptonic channel 
is reconstructed by identifying the hadron and $\ell^+$.
 The neutrino is the only missing particle, and thus the missing mass can be used to extract
signal events with a low level background. An alternative method does not tag 
one of the $D$ mesons and instead uses the entire missing energy and momentum 
in an event to access to the four momentum of the missing neutrino.
The CLEO experiment once used this method, but the large backgrounds make it
much less popular than the the double tag method.

\begin{figure}[hbpt]
\centering
\subfigure[]{\includegraphics[width=3.0in]{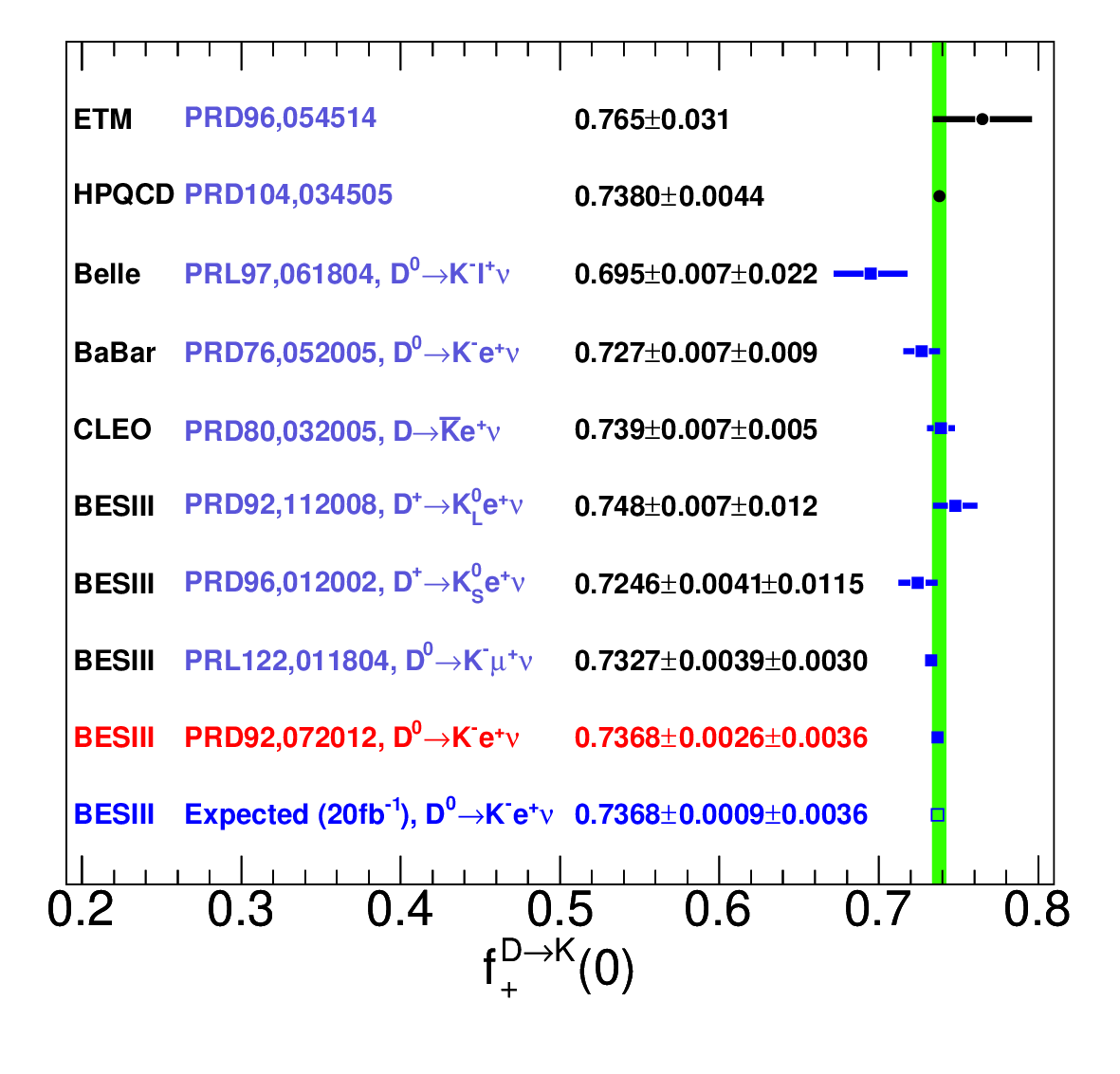}}\hskip -1pt
  \subfigure[]{\includegraphics[width=3.0in]{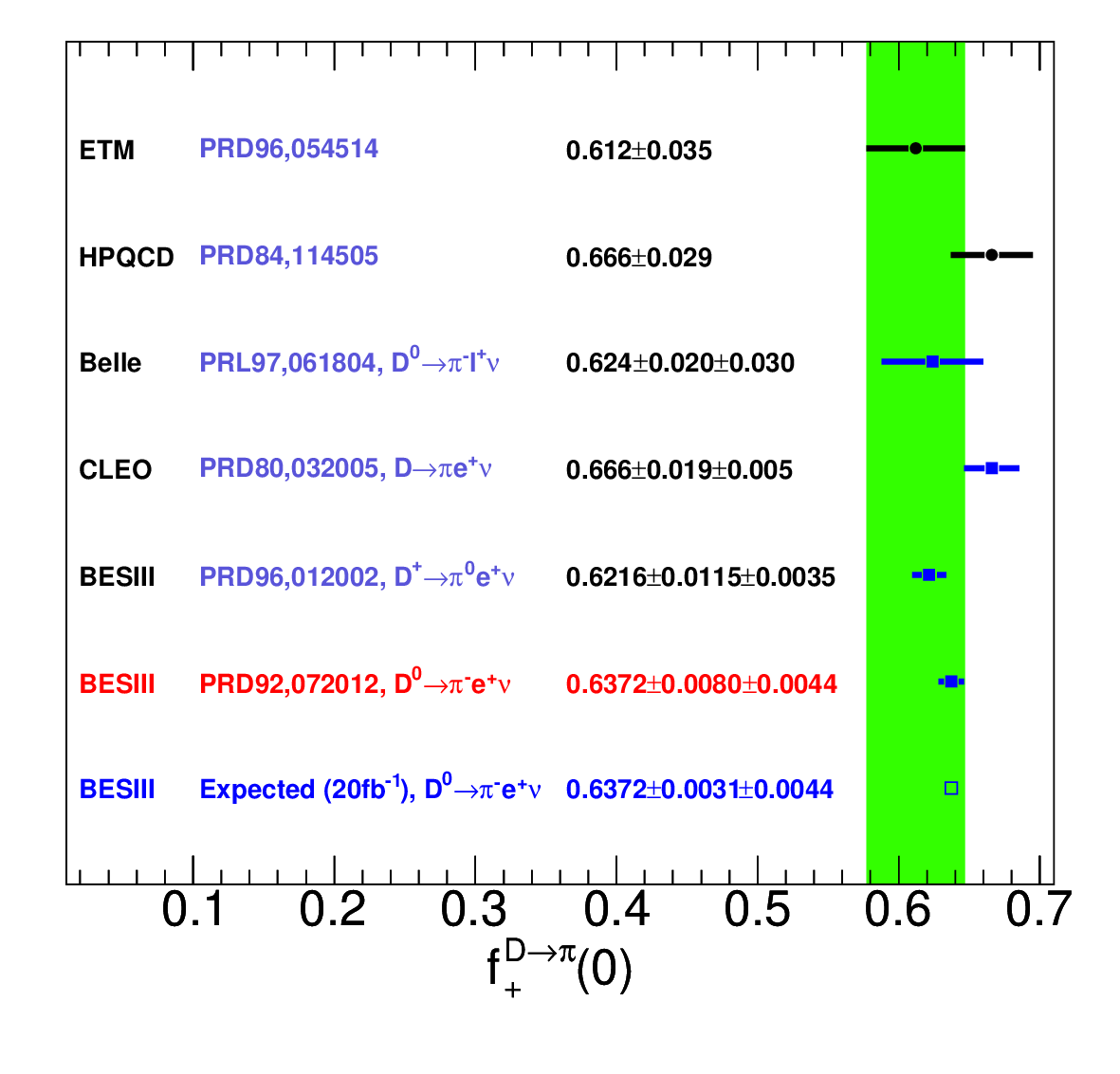}}
  \caption{Comparison of the results for (a) $f^K_+(0)$ and (b) 
  $f^\pi_+(0)$  measured  by the Belle, BaBar, CLEO-c and BESIII experiments. 
  The green bands present the lattice quantum chromodynamics~(LQCD) uncertainties. 
  Values marked in black circle denote LQCD calculations, values marked in blue
  square denotes experimental results, and the value marked in open square 
  denotes the expected precision from BESIII with 20~fb$^{-1}$ data set at 
  3.773~GeV~\cite{Ablikim:2019hff}. Note that the lattice calculation of
  $f_+^{D\to \pi}$ by HPQCD Collaboration~\cite{lqcd_fpi} was done using
  $n_f=2+1$ flavors of sea quarks, whereas HPQCD Collaboration's calculation of $f_+^{D\to K}$~\cite{Chakraborty:2021qav}
  and ETM Collaboration's calculation~\cite{lqcd_ETM} include charm quarks in the sea.
  These latter two calculations are included in the FLAG $n_f=2+1+1$ average.
  Data from Refs.~\cite{Lubicz:2017syv, Chakraborty:2021qav, Belle_D2pilnu_2006, BaBar:2007zgf, CLEO:2009svp, bes3_Dp_KLev, bes3_Dp_k0pi0ev, bes3_D0_kmuv, bes3_D0_kpiev, lqcd_fpi}. Figure provided by Hai-Long Ma.
  }
\label{fig:ff}
\end{figure}

In the past ten years, BESIII, with a triple amount of $\psi(3770)$ data,
reported measurements of these absolute decay rates and accordingly 
the products of the hadronic form factors at 
$q^2 =0$ and the magnitude of the CKM matrix elements with significantly improved precision. BESIII studied the semi-leptonic
$D_{(s)}^{0(+)}$ decays into $P$, $V$, $S$, and $A$,
where $P$ denotes pseudoscalar mesons of
$K$~\cite{bes3_D0_kpiev,bes3_Dp_k0pi0ev,bes3_Dp_KLev,bes3_D_kev,bes3_Dp_KSev_2pi0,bes3_D0_kmuv,bes3_Dp_kmuv,bes3_Dst_Kev},
$\pi$~\cite{bes3_D0_kpiev,bes3_Dp_k0pi0ev,bes3_D_pimuv},
$\eta$~\cite{bes3_Dp_etaev,bes3_Dp_etamuv,bes3_Dst_etaenu,bes3_Ds_etaev_4009,bes3_Ds_etamuv_4009},
$\eta^\prime$~\cite{bes3_Dp_etaev,bes3_Dst_etaenu,bes3_Ds_etaev_4009,bes3_Ds_etamuv_4009};
$V$ denotes vector mesons of $K^*$~\cite{bes3_Dp_kpiev,bes3_D0_kspiev},
$\rho$~\cite{bes3_D0_pipiev,bes3_D0_rhomuv},
$\omega$~\cite{bes3_Dp_omegaev,bes3_Dp_omegamuv}, and
$\phi$~\cite{bes3_Dp_omegaev,bes3_Ds_etamuv_4009};
$S$ denotes scalar mesons of $f_0$~\cite{bes3_D0_pipiev,bes3_Ds_f0enu} and
$a_0$~\cite{bes3_D_a0enu,bes3_Ds_a0enu}; and
$A$ denotes axial vector mesons of $K_1$~\cite{bes3_Dp_K1enu,bes3_D0_K1enu}
and $b_1$~\cite{bes3_D_b1enu}.
These measurements were carried out by using 2.93, 0.48, and 6.32 fb$^{-1}$ of
data taken at $\sqrt s=3.773$, 4.009, and 4.178-4.226 GeV.
From studies of the differential decay rates
(see Eq.(~\ref{eq:semi})) and the values of $|V_{cd}|$ and $|V_{cs}|$ from 
the SM-constrained fit~\cite{PDG}, one can extract the transition form factors.
The best precision of the $c\to s$ and $c\to d$ semi-leptonic $D^{0(+)}$ decay 
form factors are from the studies of $D^{0(+)}\to \bar K\ell^+\nu_\ell$ and 
$D^{0(+)}\to \pi\ell^+\nu_\ell$ (shown in Fig.~\ref{fig:ff}). 

LQCD computations of $f_+$ allow for comparisons to experiment to ascertain 
whether the SM provides the correct prediction for the $q^2$-dependence 
of $d\Gamma (D_{(s)}\to P\ell\nu_\ell)/dq^2$ and, subsequently, to determine the CKM 
matrix elements $|V_{cd}|$ and $|V_{cs}|$. As for the decay constants, we use the FLAG 
averages~\cite{FlavourLatticeAveragingGroupFLAG:2021npn} of the lattice calculations 
with $n_f=2+1+1$ sea quarks. Results included in the average come from the ETM collaboration
\cite{Lubicz:2017syv} and the HPQCD Collaboration~\cite{Chakraborty:2021qav}. Knowing the 
value of the form factor $f_+$ at $q^2=0$ is enough to determine the CKM element, and historically 
those values are usually quoted. The result for $f_+(0)$ for $D\to \bar{K}\ell\nu_\ell$
is very precise with a 0.6\% uncertainty, whereas $f_+(0)$ for $D\to \pi\ell\nu_\ell$
has an uncertainty of almost 6\%. The lattice results are shown along experimental results 
in Fig.~\ref{fig:ff} and Table~\ref{tab:semi}.

\begin{table}[htp]
 \renewcommand\arraystretch{1.25}
\centering
\caption{\label{tab:semi}
  Measurements of $D^0 /D^+$, $D^+_s$ and $\Lambda_c^+$ semi-leptonic decays with near-threshold data at BESIII and comparisons between experimental results and theoretical expectations or SM-global fit result. (Here  ``-"  indicates not available.)
For theory predictions of the form factors at $q^2=0$, we take the
Flavour Lattice Averaging Group (FLAG) averages~\cite{FlavourLatticeAveragingGroupFLAG:2021npn} of calculations with  $n_f=2+1+1$ sea quarks. The original calculations included in the average are cited next to the result. Note that these predictions, which are marked with an asterisk in the table, are for the isospin-averaged quantities. The calculation of $\Lambda_c$ form factors~\cite{Meinel:2016dqj} used $n_f=2+1$ sea quark flavors. There are currently several ongoing lattice
calculations to extract the $D_s\to K$ form factors~\cite{Marshall:2022xbz, FermilabLattice:2021bxu}. In earlier lattice calculations the $D\to\pi$ and $D_s\to K$ form factors have been found to be the same within few \% \cite{Koponen:2013ila, Koponen:2014zca}. The shape of the form factor is very insensitive to the spectator quark (here $u$ or $s$).
}
\begin{tabular}{lccc}
\hline\hline
Observable & Measurement  & Prediction/Fit \\ \hline
$\mathcal{B}(D^0 \to K^-  e^+ \nu_e) $ & $(3.505 \pm 0.014_\mathrm{stat}\pm 0.033_\mathrm{syst})\%$~\cite{bes3_D0_kpiev} &- \\
$|V_{cs}| f^{D\to K}_+ (0) $ & $0.7172 \pm 0.0025_\mathrm{stat}\pm 0.0035_\mathrm{syst}$ & - \\
$f^{D\to K}_+ (0) $ & $0.7368 \pm 0.0026_\mathrm{stat}\pm 0.0036_\mathrm{syst}$ & $0.7385 \pm 0.0044^\ast$~\cite{Lubicz:2017syv, Chakraborty:2021qav}\\ 
$\mathcal{B}(D^0 \to K^-  \mu^+ \nu_\mu) $ & $( 3.413\pm 0.019_\mathrm{stat}\pm 0.035_\mathrm{syst})\%$~\cite{bes3_D0_kmuv} &- \\
$|V_{cs}| f^{D\to K}_+ (0) $ & $0.7133 \pm 0.0038_\mathrm{stat}\pm 0.0030_\mathrm{syst}$ & - \\
$f^{D\to K}_+ (0) $ & $0.7327 \pm 0.0039_\mathrm{stat}\pm 0.0030_\mathrm{syst}$ & $0.7385 \pm 0.0044^\ast$~\cite{Lubicz:2017syv, Chakraborty:2021qav}\\ 
$\mathcal{B}(D^0 \to K^-  \mu^+ \nu_\mu) $/$\mathcal{B}(D^0 \to K^-  e^+ \nu_e) $ & $0.974 \pm 0.007stat_\mathrm{stat}\pm 0.012_\mathrm{syst}$~\cite{bes3_D0_kmuv}& $0.9779 \pm 0.0050$~\cite{Chakraborty:2021qav} \\
\hline
$\mathcal{B}(D^0 \to \pi^-  e^+ \nu_e)$ &$(0.295 \pm 0.004_\mathrm{stat}\pm 0.003_\mathrm{syst})\%$~\cite{bes3_D0_kpiev}  &- \\ \
$|V_{cd}| f^{D\to \pi}_+ (0) $ & $ 0.1435\pm 0.0018_\mathrm{stat}\pm 0.0009_\mathrm{syst}$  & - \\
$f^{D\to \pi}_+ (0) $ & $0.6372 \pm 0.0080_\mathrm{stat} \pm 0.0044_\mathrm{syst}$  &  $0.612\pm 0.035^\ast$~\cite{Lubicz:2017syv} \\ \hline
$\mathcal{B}(D^+ \to \bar{K}^0 e^+\nu_e) $ & $ (8.60 \pm 0.06_\mathrm{stat} \pm 0.15_\mathrm{syst})\%$~\cite{bes3_Dp_k0pi0ev} & - \\
$f^{D\to K}_+ (0) $ & $  0.725 \pm 0.004_\mathrm{stat}\pm 0.012_\mathrm{syst}$ & $0.7385 \pm 0.0044^\ast$~\cite{Lubicz:2017syv, Chakraborty:2021qav}\\ \hline
$\mathcal{B}(D^+ \to \pi^0  e^+\nu_e ) $ & $ (0.363\pm 0.008_\mathrm{stat} \pm 0.005_\mathrm{syst} )\%$~\cite{bes3_Dp_k0pi0ev} & - \\
$f^{D\to \pi}_+ (0)  $ & $ 0.622 \pm 0.012_\mathrm{stat} \pm 0.003_\mathrm{syst}$  & $0.612\pm 0.035^\ast$~\cite{Lubicz:2017syv} \\ \hline
 $f^{D\to \pi}_+ (0)/f^{D\to K}_+ (0) $ & $0.865\pm 0.013$~\cite{bes3_Dp_k0pi0ev}  & $0.84\pm 0.04$~\cite{Ball:2006yd} \\ \hline
 
 $\mathcal{B}(D^+ \to \eta  e^+ \nu_e)$ &$(10.74 \pm 0.81_\mathrm{stat}\pm 0.51_\mathrm{syst})\times10^{-4}$~\cite{bes3_Dp_etaev}  &- \\ 
$|V_{cd}| f^{D\to \eta}_+ (0) $ & $ 0.0079\pm 0.0006_\mathrm{stat}\pm 0.0002_\mathrm{syst}$  & - \\
  $\mathcal{B}(D^+ \to \eta  \mu^+ \nu_\mu)$ &$(10.4 \pm 1.0_\mathrm{stat}\pm 0.5_\mathrm{syst})\times10^{-4}$~\cite{bes3_Dp_etamuv}  &- \\
$|V_{cd}| f^{D\to \eta}_+ (0) $ & $ 0.0087\pm 0.0008_\mathrm{stat}\pm 0.0002_\mathrm{syst}$  & - \\
 $f^{D\to\eta}_+ (0) $ & $0.39 \pm 0.04_\mathrm{stat} \pm 0.01_\mathrm{syst}$  &- \\ 
 $\mathcal{B}(D^+ \to \eta  \mu^+ \nu_\mu)$/$\mathcal{B}(D^+ \to \eta e^+ \nu_e)$ & $0.91\pm 0.13$&-\\ 
 \hline
$\mathcal{B}(D_s^+ \to \eta  e^+ \nu_e)$ &$(2.323 \pm 0.063_\mathrm{stat}\pm 0.063_\mathrm{syst})\%$~\cite{bes3_Dst_etaenu}  & - \\ \
$|V_{cs}| f^{D_s\to \eta}_+ (0) $ & $ 0.4455\pm 0.0053_\mathrm{stat}\pm 0.0044_\mathrm{syst}$  & - \\
 $f^{D_s\to\eta}_+ (0) $ & $0.4576 \pm 0.0054_\mathrm{stat} \pm 0.0045_\mathrm{syst}$  &  - \\\hline
 $\mathcal{B}(D_s^+ \to \eta^{\prime}  e^+ \nu_e)$ &$(2.323 \pm 0.063_\mathrm{stat}\pm 0.063_\mathrm{syst})\%$~\cite{bes3_Dst_etaenu}  &- \\ \
$|V_{cs}| f^{D_s\to\eta^{\prime}}_+ (0) $ & $ 0.477\pm 0.049_\mathrm{stat}\pm 0.011_\mathrm{syst}$  & - \\
 $f^{D_s\to\eta^{\prime}}_+ (0) $ & $0.490 \pm 0.050_\mathrm{stat} \pm 0.011_\mathrm{syst}$  &  - \\\hline
 $\mathcal{B}(D_s^+ \to K^0 e^+ \nu_e)$ &$(3.25 \pm 0.38_\mathrm{stat}\pm 0.16_\mathrm{syst})\%$~\cite{bes3_Dst_Kev}  &- \\ \
$|V_{cd}| f^{D_s\to K}_+ (0) $ & $ 0.162\pm 0.019_\mathrm{stat}\pm 0.003_\mathrm{syst}$  & - \\
 $f^{D_s\to K}_+ (0) $ & $0.720 \pm 0.084_\mathrm{stat} \pm 0.013_\mathrm{syst}$  & $f^{D_s\to K}(0)\approx f^{D\to\pi}(0)$ within 5\% \cite{Koponen:2013ila, Koponen:2014zca}\\
\hline
$\mathcal{B}(\Lambda^+_c \to \Lambda e^+ \nu_e )$ & $(3.56 \pm 0.11_\mathrm{stat} \pm 0.07_\mathrm{syst})\%$~\cite{2022_Lamdbac_Lenu} & $3.55 \pm 1.04\%$~\cite{LightFront_CharmBaryonSemi} \\
 $\alpha_{1}^{g_{\perp}}$ & $1.43 \pm 2.09_\mathrm{stat} \pm 0.16_\mathrm{syst}$ & - \\
 $\alpha_{1}^{f_{\perp}}$ & $-8.15 \pm 1.58_\mathrm{stat} \pm 0.05_\mathrm{syst}$ & - \\
  $r_{f_{+}}$ & $1.75 \pm 0.32_\mathrm{stat} \pm 0.01_\mathrm{syst}$ & - \\
   $r_{f_{\perp}}$ & $3.62 \pm 0.65_\mathrm{stat} \pm 0.02_\mathrm{syst}$ & - \\
  $r_{g_{+}}$ & $1.13 \pm 0.13_\mathrm{stat} \pm 0.01_\mathrm{syst}$ & - \\  
 $\mathcal{B}(\Lambda^+_c \to \Lambda \mu^+ \nu_\mu)$ &  $(3.49\pm 0.46_\mathrm{stat} \pm 0.27_\mathrm{syst})\%$~\cite{Ablikim:2016vqd}  & - \\
 $\mathcal{B}(\Lambda^+_c \to  \Lambda \mu^+ \nu_\mu)/\mathcal{B}(\Lambda^+_c \to  \Lambda e^+ \nu_e) $ & $0.96\pm 0.16_\mathrm{stat} \pm 0.04_\mathrm{syst}$ & $\approx 1.0$~\cite{Angular_LFU_CharmBaryon} \\
 $\Gamma(\Lambda_c \to \Lambda e^+ \nu_e)/|V_{cs}|^2$ & - & $0.2007 \pm 0.0071_\mathrm{stat} \pm 0.0074_\mathrm{syst}$ ps$^{-1}$~\cite{Meinel:2016dqj} \\
 $\Gamma(\Lambda_c \to \Lambda \mu^+ \nu_\mu)/|V_{cs}|^2$ & - & $0.1945 \pm 0.0069_\mathrm{stat} \pm 0.0072_\mathrm{syst}$ ps$^{-1}$~\cite{Meinel:2016dqj} \\
\hline\hline
\end{tabular}
\end{table}

The most commonly used parameterization of the form factors is the $z$-expansion:
\begin{equation}
f(q^2)=\frac{1}{B(q^2)\phi(q^2,t_0)}\sum_{n=0}^{N}a_n z^n,\qquad
z(q^2,t_0)=\frac{\sqrt{t_+-q^2}-\sqrt{t_+-t_0}}{\sqrt{t_+-q^2}+\sqrt{t_+-t_0}},
\end{equation}
which exploits the positivity and analyticity properties of two-point functions of 
vector currents. This is a conformal transformation that maps the $q^2$ plane cut for 
$q^2\ge t_+$ onto the disk $|z(q^2,t_0)|<1$ in the $z$ complex plane. 
Here $t_+=(M_{D_{(s)}}+M_P)^2$ (the meson $P$ being $\pi$ or $K$), and the real 
parameter $t_0$ can be chosen freely as long as $t_0<t_+$. A common choice is
$t_0=t_+(1-(1-t_-/t_+)^{1/2})$, which minimises the maximum value of $z$ over
the $q^2$ range of the decay. $B(q^2)$ is the
 Blaschke factor, which contains poles and cuts below $t_+$. Common
choices are $B(q^2)=1-q^2/M^2_{\mathrm{pole}}$, where $M_{\mathrm{pole}}$ is the 
pole mass, or $B(q^2)=z(q^2,M^2_{\mathrm{pole}})$. The so-called outer function
$\phi(q^2,t_0)$ can have a more complicated $q^2$ dependence, or it can be chosen
to be $\phi(q^2,t_0)=1$. 

\begin{figure}[hbpt]
\centering
\includegraphics[width=0.52\textwidth]{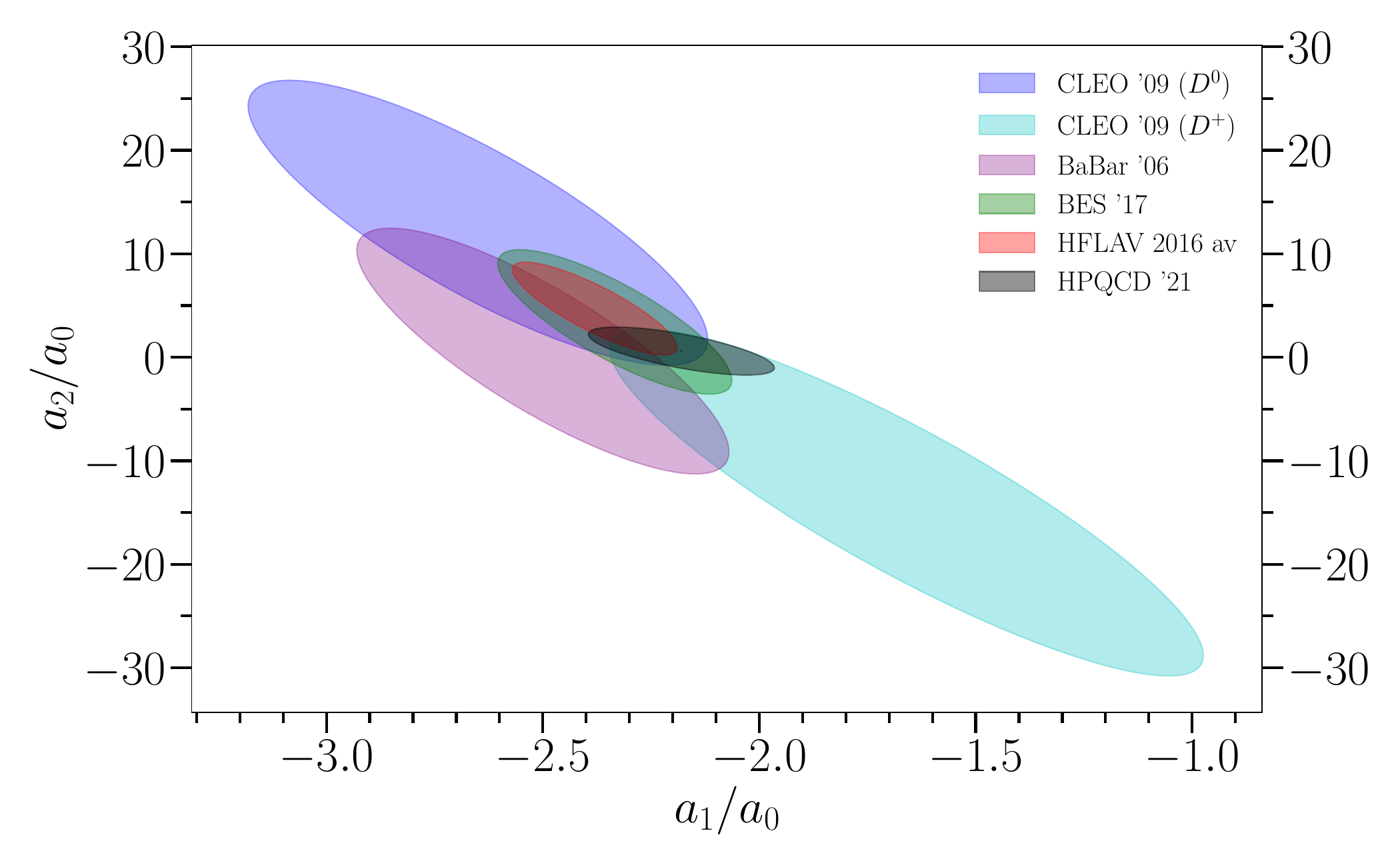}
  \caption{Comparison of the shape of the vector form factor
    for $D \to \bar{K}\ell\nu_\ell$ expressed in terms of ratios of the z-expansion coefficients.
    Ellipses give the 68\% confidence limits ($\Delta\chi^2 = 2.3$). Experimental
    results are from \cite{CLEO:2009svp, BaBar:2007zgf, bes3_D0_kpiev,HFLAV:2016hnz}.
    CLEO results are for $D^0 \to K^- e^+ \nu_e$
    (dark blue) and $D^+ \to \bar{K}^0 e^+\nu_e$ (light blue); all other experimental
    data is for $D^0 \to K^- e^+\nu_e$. The HFLAV experimental average~\cite{HFLAV:2016hnz}
    is given as the red ellipse. The black ellipse gives the LQCD result by
    the HPQCD Collaboration \cite{Chakraborty:2021qav}. Good agreement is seen between the
    experimental results and the LQCD result. Figure from~\cite{Chakraborty:2021qav}.
}
\label{fig:errorellipse}
\end{figure}

If one uses the same $z$-expansion to fit lattice form
factors and different experimental data sets, or uses a 'refitting' procedure as
in~\cite{Chakraborty:2021qav}, one can compare the shapes of the form factors
extracted from each data set by looking at ratios of the coefficients $a_n$ of the
$z$-expansion. (See~\cite{Chakraborty:2021qav} Eq. B2 for the outer function $\phi(q^2,t_0)$
used there.) Since the CKM element cancels in the ratio, the comparison is direct.
This is shown in Fig.~\ref{fig:errorellipse} (which is from~\cite{Chakraborty:2021qav}),
comparing form factors extracted from CLEO~\cite{CLEO:2009svp}, BaBar~\cite{BaBar:2007zgf}, 
BESIII~\cite{bes3_D0_kpiev} as well as the HFLAV experimental average~\cite{HFLAV:2016hnz} 
and the LQCD determination by HPQCD~\cite{Chakraborty:2021qav}. The agreement is
seen to be good, and possible tensions in the shape of the form factor between theory and 
experiment or between different experiments would be apparent in a comparison like this one.

\begin{figure}[hbpt]
\centering
\includegraphics[width=0.472\textwidth]{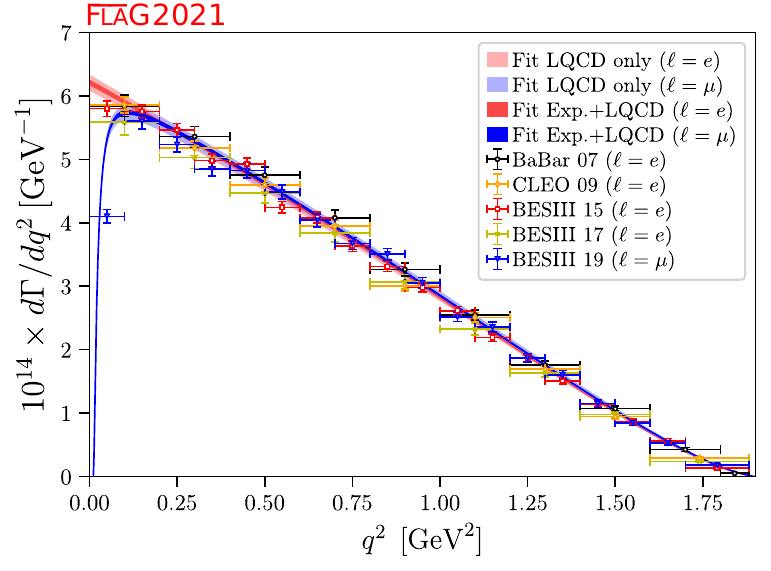}
  \caption{\small The $D\to \bar{K}\ell\nu_\ell$ differential decay rates from the FLAG 2021
    review~\cite{FlavourLatticeAveragingGroupFLAG:2021npn}. This illustrates the FLAG fit to
    LQCD data and to LQCD+experimental data (see section 7 of the review for further details on the fits).}
\label{fig:fitLatplusExpt}
\end{figure}

The FLAG review~\cite{FlavourLatticeAveragingGroupFLAG:2021npn} also provides $z$-expansion
parameterizations of the form factors for the $D\to \bar{K}\ell\nu_\ell$ decay. They adopt the 
standard implementation of the
Bourrely-Caprini-Lellouch (BCL) parameterization~\cite{Bourrely:2008za}, which is
widely used by experimental and LQCD collaborations:
\begin{align}
f_+(q^2)&=\frac{1}{1-q^2/(M^V_{\mathrm{pole}})^2}\sum_{n=0}^{N-1}a^+_n\left[z^n-(-1)^{n-N}\frac{n}{N}z^N\right],\nonumber \\
f_0(q^2)&=\frac{1}{1-q^2/(M^S_{\mathrm{pole}})^2}\sum_{n=0}^{N-1}a^0_n z^n,
\end{align}
where $M^V_{\mathrm{pole}}$ is the vector pole mass ($D^\ast_s$ mass in the case
of $D\to \bar{K}\ell\nu_\ell$ decay), and $M^S_{\mathrm{pole}}$ is the scalar pole mass ($D_{s0}$  
mass for charm-to-strange decay). This is illustrated in Fig.~\ref{fig:fitLatplusExpt},
which shows FLAG's fit to lattice data only (normalizing the fit
result for the plot with their choice of $|V_{cs}|$) and also their combined fit to both lattice
and experimental data (in which case $|V_{cs}|$ can be extracted from the fit).

The FLAG $n_f=2+1+1$ averages~\cite{FlavourLatticeAveragingGroupFLAG:2021npn} for the CKM
elements from these fits are~\cite{Lubicz:2017syv, Riggio:2017zwh, Chakraborty:2021qav}
\begin{align}
    |V_{cd}| =& 0.2341\pm74\quad\\
    |V_{cs}| =& 0.9714\pm69\quad.
\end{align}
The experimental data sets included in the fit for $|V_{cd}|$ are $D^0\to \pi^-e^+\nu_e$ from
BaBar~\cite{BaBar:2014xzf}, 
$D^0\to \pi^-e^+\nu_e$ and $D^+\to \pi^0e^+\nu_e$ from CLEO-c~\cite{CLEO:2009svp} ,
$D^0\to \pi^-e^+\nu_e$ and $D^+\to \pi^0e^+\nu_e$ from and BESIII~\cite{bes3_D0_kpiev,bes3_Dp_k0pi0ev}.
In the extraction of $|V_{cs}|$ the experimental data consist of $D^0\to K^-e^+\nu_e$ from
BaBar~\cite{BaBar:2007zgf}, CLEO-c~\cite{CLEO:2009svp} and BESIII~\cite{bes3_D0_kpiev}; 
$D^+\to \bar{K}^0e^+\nu_e$ from CLEO-c~\cite{CLEO:2009svp} and BESIII~\cite{bes3_Dp_k0pi0ev}; 
and  $D^0\to K^-\mu^+\nu_\mu$ from BESIII~\cite{bes3_D0_kmuv}.
Given the form factor $f_+(q^2)$ from LQCD, one can
integrate (numerically) Eq.~\eqref{eq:semi} over $q^2$ bins and extract the CKM element
bin-by-bin basis by taking the ratio with the experimental value for that bin. Any
tensions between the shapes of the form factors from different data sets would again
be distinct. 

To study the $c\to d$ transition, one can also look at the $D^+_s\to K^0e^+\nu_e$ decay, in addition
to $D^0\to \pi^-e^+\nu_e$ and $D^+\to \pi^0e^+\nu_e$. The difference between these decays is the spectator quark,
which is here either $s$ or $u$ quark. There are currently several ongoing lattice
calculations to extract the $D_s\to K\ell\nu$ form factors~\cite{Marshall:2022xbz, FermilabLattice:2021bxu}. 
In earlier lattice calculations the $D\to\pi\ell\nu$ and $D_s\to K\ell\nu$ form factors have been 
found to be the same within few \% in the whole $q^2$ range allowed by 
kinematics~\cite{Koponen:2013ila, Koponen:2014zca}. The shape of the form 
factor is very insensitive to the spectator quark.

In addition to $D_{(s)}$ meson semileptonic decays to a single pseudoscalar meson (plus the lepton and 
its neutrino),  BESIII has reported the partial wave analyses of 
$D^{0(+)}\to \pi^{0(+)}\pi^{-}e^{+}\nu_e$~\cite{bes3_D0_pipiev} and 
$D^{+} \to K^{-} \pi^+ e^+\nu_e$~\cite{bes3_Dp_kpiev,bes3_D0_kspiev}, and 
measures the parameters of the corresponding form factors. BESIII also has observed
the $D^{0(+)}$ to axial-vector particle semileptonic decays, 
$D^{0(+)}\to K_1(1270)^{-(0)}e^{+}\nu_e$~\cite{bes3_Dp_K1enu,bes3_D0_K1enu}, and 
$D^{0(+)}$ to scalar particle semileptonic decays, 
$D^{0(+)}\to a_0(980)^{-(0)}e^{+}\nu_e$~\cite{bes3_D_a0enu} for the first 
time. These studies on the intermediate resonances in hadronic final states, 
e.g., $K_1(1270)$, $K^{*}(980)$, $f_0(980)$, and $a_0(980)$, in the semileptonic 
$D^{0(+)}_{(s)}$ decays provide a clean environment to explore meson spectroscopy, 
as no other particles interfere. 
This is a major advantage in comparison to hadronic spectroscopy studies 
in charmonium decay or hadronic $D_{(s)}$ decays.

The $D_s^+$ to vector meson semi-leptonic decay $D_s^+\to \phi\ell^+\nu_\ell$
offers another independent approach to $|V_{cs}|$. The first experimental measurement 
is reported by Babar~\cite{BaBar:2008gpr}. On the theory side, there is a LQCD calculation 
of the decay~\cite{Donald:2013pea}, which is more complicated as there are more form 
factors involved. This calculation was done using $n_f=2+1$ sea quarks. 
 In~\cite{Donald:2013pea} the authors also extracted the $q^2$ and angular distributions 
 for the differecitential rate and found good agreement with those
from the BaBar experiment. From the total branching fraction they obtain $|V_{cs}| = 1.017(63)$,
in good agreement with that from $D\to \bar{K}\ell\nu_\ell$ semileptonic decay. 

\begin{figure}[hbpt]
  \centering
  \subfigure[]{\includegraphics[height=2.5in]{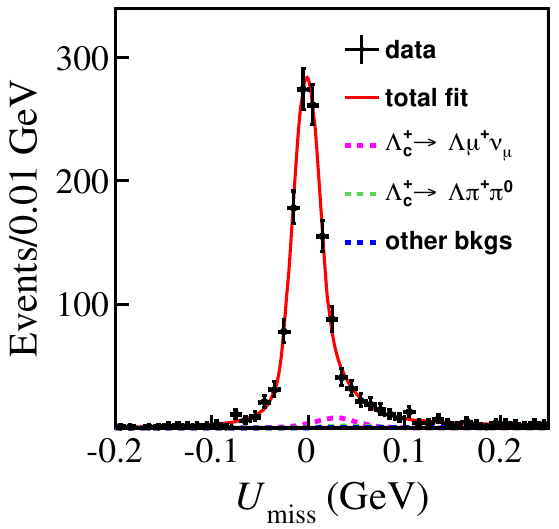}}\hskip -1pt
  \subfigure[]{\includegraphics[height=2.5in]{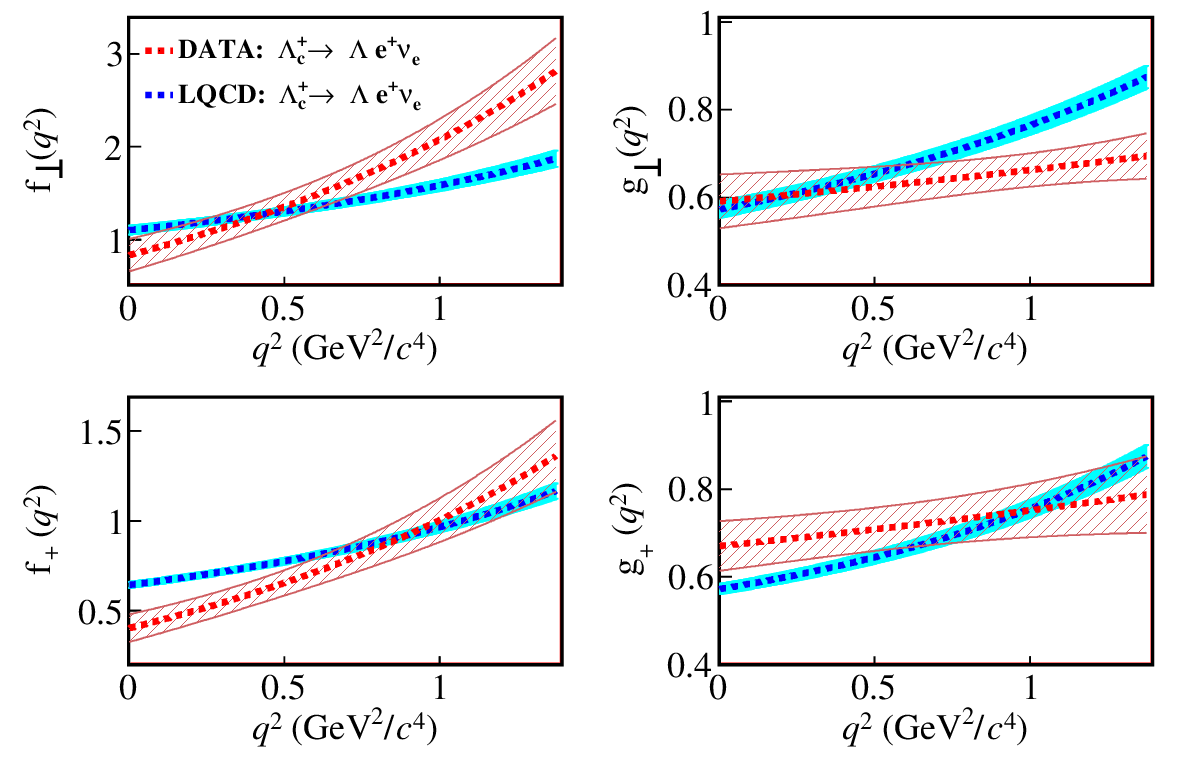}}
  \caption{(a) Fit to the $U_{\rm miss}$ distribution within the
    $\Lambda$ signal region~\cite{2022_Lamdbac_Lenu}. The points with error
    bars are data, the solid curves show the total fits, and the dashed curves are the background shapes. 
    (b) Comparison of the form factors with a lattice quantum chromodynamics~(LQCD)
    calculation~\cite{Meinel:2016dqj}. The bands show the total uncertainties.
    Figure from Ref.~\cite{2022_Lamdbac_Lenu}.
  }
  \label{fig:Lc_signals-2}
\end{figure}

A similar approach can be applied to the charmed baryon semi-leptonic decays.
Among the $\Lambda_c^{+}$ semileptonic decays, $\Lambda_c^{+}\to\Lambda \ell^+\nu_\ell$
is the only dominant one. BESIII, using the similar tag technique employed in the 
$D\bar{D}$ threshold production, measures the absolute branching fraction of 
$\Lambda_c^{+}\to\Lambda e^+\nu_e$ and the $\Lambda_c^{+}\to \Lambda$ form 
factors for the first time based on 4.5~fb$^{-1}$ $\Lambda_c^{+}\bar{\Lambda}_c^{-}$ dataset collected
at $4.600$-$4.699$~GeV~\cite{Ablikim:2015prg, 2022_Lamdbac_Lenu}. 
The $\bar{\Lambda}_c^-$ is reconstructed via its hadronic decay modes and the missing neutrino on the signal 
side is inferred by the kinematic variable $U_{\rm miss} = E_{\rm miss}-c|\vec{p}_{\rm miss}|$, where $E_{\rm miss}$ and 
$\vec{p}_{\rm miss}$ are the missing energy and missing momentum, respectively. 
The $U_{\rm miss}$ distribution is presented in Fig.~\ref{fig:Lc_signals-2}. The measured branching
fractions of $\Lambda_c^{+}\to\Lambda e^+\nu_e$ and $\Lambda_c^{+}\to\Lambda \mu^+\nu_\mu$, and their ratio are given in 
Table~\ref{tab:semi}. 
Four form factors $f_{\bot,+}(q^2)$ and $g_{\bot,+}(q^2)$ are necessary to parameterize 
the dynamics, defined following a 
$z$-expansion:  
\begin{eqnarray}
f(q^2)&=&\frac{a_0^{f}}{1-q^2/\left(m_{\rm pole}^{f}\right)^2} \left[1+\alpha^{f}_1\times z(q^2)\right],
\label{eq:formfactor}
\end{eqnarray}
where $m^f_{\rm pole}$ is the pole mass; $a_0^f$, $\alpha_1^f$ are free parameters; 
$z(q^2)=\frac{(\sqrt{t_+-q^2}-\sqrt{t_+-t_0})}{(\sqrt{t_+-q^2}+\sqrt{t_+-t_0})}$ with $t_0=q^2_{\rm max}=(m_{\Lambda_c}-m_{\Lambda})^2$, $t_+=(m_D+m_K)^2$, $m_D=1.870$~GeV/$c^2$ and $m_K=0.494$~GeV/$c^2$. 
The pole masses $m_{\rm pole}^{f_+,f_{\perp}}$ and $m_{\rm pole}^{g_+,g_{\perp}}$ are fixed to 2.112 and 2.460~GeV/$c^2$, respectively~\cite{Meinel:2016dqj}.
Setting $\alpha^{g_{\perp}}_1\equiv \alpha^{g_{+}}_1$ and $\alpha^{f_{\perp}}_1\equiv \alpha^{f_{+}}_1$,
BESIII measures
$\alpha^{g_{\perp}}_1$, $\alpha^{f_{\perp}}_1$, $r_{f_{+}}=a_0^{f_{+}}/a_0^{g_{\perp}}$, $r_{f_{\perp}}=a_0^{f_{\perp}}/a_0^{g_{\perp}}$, and $r_{g_{+}}=a_0^{g_{+}}/a_0^{g_{\perp}}$,
given in Table~\ref{tab:semi}. In addition, BESIII reports the observation of 
$\Lambda_c^{+}\to pK^{-} e^+\nu_e$ and evidence for 
$\Lambda_c^{+}\to \Lambda^{*}(1520) e^+\nu_e$~\cite{2022_L1520enu}.

There exists one LQCD calculation of the $\Lambda_c^+ \to \Lambda \ell^+ \nu_{\ell}$ form factors 
and decay rates by Meinel~\cite{Meinel:2016dqj}. The main results of the paper are the LQCD
predictions for the decay rates:
$\Gamma(\Lambda_c^+ \to \Lambda e^+ \nu_e)/|V_{cs}|^2 = 0.2007\pm71\pm74$~ps$^{-1}$
and $\Gamma(\Lambda_c^+ \to \Lambda \mu^+ \nu_\mu)/|V_{cs}|^2 = 0.1945\pm69\pm72$~ps$^{-1}$,
where the first uncertainties are statistical and the second ones are systematic
uncertainties. Combining their results with BESIII measurements of the branching
fractions~\cite{Ablikim:2015prg,Ablikim:2016vqd} and using $\tau_{\Lambda_c}= 0.200\pm6$~ps 
they extract the CKM element
\begin{equation}
 |V_{cs}| = \left\{
 \begin{array}{ll} 0.951\pm24_{\rm LQCD}\pm14_{\tau_{\Lambda_c}}\pm56_{\mathcal{B}}, & \ell=e, \\
0.947\pm24_{\rm LQCD}\pm14_{\tau_{\Lambda_c}}\pm72_{\mathcal{B}}, & \ell=\mu, \\
0.949\pm24_{\rm LQCD}\pm14_{\tau_{\Lambda_c}}\pm49_{\mathcal{B}}, & \ell=e,\mu, \end{array}\right.
\end{equation}
where the last line is the correlated average over $\ell=e,\mu$.
The LQCD form factors and the new BESIII results~\cite{2022_Lamdbac_Lenu} are
compared in Fig.~\ref{fig:Lc_signals-2} (b). Some tensions can be observed
between the shapes of the form factors, but smaller uncertainties are needed to
draw any conclusions.

In addition, there is a LQCD  determination of the $\Lambda_c^+\to N$ (where
$N$ is a proton or a neutron) vector, axial vector, and tensor
form factors~\cite{Meinel:2017ggx}. The results provide SM predictions for 
the $\Lambda_c^+\to ne^+\nu_e$ and $\Lambda_c^+\to n\mu^+\nu_\mu$ decay rates
with an uncertainty of 6.4\%. These modes can be only studied with double-tag
method which is unique at BESIII, and they are in the list of the BESIII charmed baryon project.  
The form factors can also be applied to study the differential branching fraction
and angular distribution of the rare charm decay $\Lambda_c^+\to p\mu^+\mu^-$,
using perturbative results for the effective Wilson coefficients in the SM.


Besides $\Lambda_c^+$, there are also studies for semileptonic decays of charmed 
baryon $\Xi_c^{+,0}$ and $\Omega_c^{0}$. The $\Xi_c^{+,0}\to \Xi^{0,-}$, like 
$\Lambda_c^{+}\to \Lambda$, are $1/2^+\to1/2^+$ transitions, which allows a 
relatively simple theoretical calculation of form factors and hadronic 
structures under nonperturbative QCD. 
The authors of \cite{lqcd_Xic2Xilnu_2021} recently performed the first LQCD
calculation of the $\Xi_c\to \Xi\ell\nu_\ell$ form factors. This calculation
uses two lattice ensembles with different lattice spacings and $n_f=2+1$ sea
quark flavours. The pion masses on both ensembles are unphysical and nearly
identical, 290~MeV and 300~MeV. The results are
extrapolated to the continuum limit, but are not extrapolated to the physical pion mass.
Therefore the error budget is not complete, as systematic uncertainty for the effect of the missing
chiral extrapolation is not estimated. However, this is an important first step
and shows that LQCD calculations of $\Xi_c$ form factors are now feasible.

For the $\Omega_c^0$ weak decays, most decay channels involve non-factorizable 
effects. Nonetheless, $\Omega_c^0\to \Omega^- M^+$ with $M=(\pi,\rho)$ and 
$\Omega_c^0\to \Omega^- \ell^+\nu_\ell$ with $\ell=(e,\mu)$ proceed through the 
$\Omega_c^0\to \Omega^-$, which is a $1/2^+\to 3/2^+$ transition, along with
the external $W$-boson emitted to connect $M^+$ or $\ell^+\nu_\ell$. There exist 
eight form factors for the $\Omega_c^0\to \Omega^-$ transition, two more than that 
in $1/2^+\to1/2^+$ transitions, which play the key role in calculating
${\cal B}(\Omega_c^0\to\Omega^-M^+)$ and 
${\cal B}(\Omega_c^0\to\Omega^-e^+\nu_e)$~\cite{LightFront_Qarkmodel_Omegac,Xu:1992sw,Cheng:1996cs,Gutsche:2018utw,Pervin:2006ie}.
Belle, CLEO II, and ALICE have performed 
the most precise branching fraction measurement of 
$\Xi_c^{+,0}\to \Xi^{0,-}\ell^+\nu_\ell$~\cite{belle_Xic_Xilnu, ALICE_Xic_Xienu, CLEO_Xic_Xienu},
and the branching ratios 
${\cal B}(\Omega_c^0\to\Omega^-e^+\nu_e)/{\cal B}(\Omega_c^0\to\Omega^-\pi^+)$~\cite{CLEO_Omegac_Omegaenu, Belle_Omegac_Omegalnu} while the 
experimental results are not well consistent with the theoretical predictions. 
The relevant studies are summarized in Table~\ref{tab:Xic_semi}.

\begin{table}[htp]
  \renewcommand\arraystretch{1.25}
  \centering
  \caption{\label{tab:Xic_semi}
    Measurements of $\Xi_c$, $\Omega_c$ semi-leptonic decays and comparisons between experimental results and theoretical expectations.
    The theoretical works of LQCD, LCSR, SU(3), LFQM, RQM, and QCDSR are compared.
    Abbreviations: LCSR, light-cone QCD sum rules; LFQM, light-front quark model; LQCD, lattice QCD.
    (Here  ``-"  indicates not available.)}
  \begin{tabular}{lccc}
    \hline\hline
    Observable/Measurement & experiment & Prediction/Fit \\ \hline
    \multirow{11}{*}{\shortstack[l]{$\mathcal{B}(\Xi^0_c \to \Xi^- e^+ \nu_e )$\\=$(1.31 \pm 0.04_\mathrm{stat} \pm 0.07_\mathrm{syst} \pm 0.38_\mathrm{\Xi^0_c \to \Xi^- \pi^+})\%$~\cite{belle_Xic_Xilnu}}} & \multirow{11}{*}{Belle} &  $2.38 \pm 0.30_\mathrm{stat} \pm 0.32_\mathrm{syst} \%$ ~\cite{lqcd_Xic2Xilnu_2021} (LQCD) \\ 
    & & $4.10 \pm 0.46 \%$ ~\cite{SU3_Xic2Xlnu_2021} (SU(3)) \\
    & & $(3.0 \pm 0.3, 2.4 \pm 0.3, 2.7 \pm 0.2) \%$ ~\cite{SU3_Xic2Xlnu_Geng2019} (SU(3)) \\
    & & $3.4 \pm 1.7 \%$ ~\cite{qcdsum_Xic2Xilnu} (QCDSR) \\
    & & $3.49 \pm 0.95 \%$ ~\cite{LightFront_CharmBaryonSemi} (LFQM) \\
    & & $1.72 \pm 0.35 \%$ ~\cite{LightFront_Xic2Xienu} (LFQM) \\
    & & $7.26 \pm 2.54 \%$ ~\cite{LightConeQCD_Xic2XiSemi} (LCSR) \\
    & & $1.85 \pm 0.56 \%$ ~\cite{LCSR_Xic2Xilnu_2021} (LCSR) \\
    & & $2.81^{+0.17}_{-0.15} \%$ ~\cite{LCSR_Xic2Xilnu_Huang} (LCSR) \\
    & & $1.35\%$ ~\cite{Lightfront_Xic2Xilnu_Zhao} (LFQM) \\
    & & $2.38\%$ ~\cite{relativistitc_Xic2Xilnu} (RQM) \\
    \hline
    \multirow{7}{*}{\shortstack[l]{$\mathcal{B}(\Xi^0_c \to \Xi^- \mu^+ \nu_\mu )$\\=$(1.27 \pm 0.06_\mathrm{stat} \pm 0.10_\mathrm{syst} \pm 0.37_\mathrm{\Xi^0_c \to \Xi^- \pi^+})\%$~\cite{belle_Xic_Xilnu}}} & \multirow{7}{*}{Belle} & $2.29 \pm 0.29_\mathrm{stat} \pm 0.31_\mathrm{syst} \%$ ~\cite{lqcd_Xic2Xilnu_2021} (LQCD) \\
    & & $3.98 \pm 0.57 \%$ ~\cite{SU3_Xic2Xlnu_2021} (SU(3)) \\
    & & $3.34 \pm 0.94 \%$ ~\cite{LightFront_CharmBaryonSemi} (LFQM) \\
    & & $7.15 \pm 2.50 \%$ ~\cite{LightConeQCD_Xic2XiSemi} (LCSR) \\
    & & $1.79 \pm 0.54 \%$ ~\cite{LCSR_Xic2Xilnu_2021} (LCSR) \\
    & & $2.72^{+0.17}_{-0.15} \%$ ~\cite{LCSR_Xic2Xilnu_Huang} (LCSR) \\
    & & $2.31\%$ ~\cite{relativistitc_Xic2Xilnu} (RQM) \\
    $\mathcal{B}(\Xi^0_c \to  \Xi^- e^+ \nu_e)/\mathcal{B}(\Xi^0_c \to \Xi^- \mu^+ \nu_\mu) $ & \multirow{2}{*}{Belle} & 
    \multirow{2}{*}{$\approx 1.0$~\cite{Angular_LFU_CharmBaryon}} \\
    =$1.03\pm 0.05_\mathrm{stat} \pm 0.07_\mathrm{syst}$~\cite{belle_Xic_Xilnu}& &\\
    \hline
    $\mathcal{B}(\Xi^0_c \to \Xi^- e^+ \nu_e )/\mathcal{B}(\Xi^0_c \to \Xi^- \pi^+ )$ &\multirow{2}{*}{ALICE} & \multirow{2}{*}{-} \\
    =$1.38 \pm 0.14_\mathrm{stat} \pm 0.22_\mathrm{syst}$~\cite{ALICE_Xic_Xienu} & &  \\
    \hline
    \multirow{9}{*}{\shortstack[l]{$\mathcal{B}(\Xi^+_c \to \Xi^0 e^+ \nu_e )$\\=$(7 \pm 4)\%$~\cite{PDG,CLEO_Xic_Xienu}} }& \multirow{9}{*}{CLEO II} & $7.18 \pm 0.90_\mathrm{stat} \pm 0.98_\mathrm{syst} \%$ ~\cite{lqcd_Xic2Xilnu_2021} (LQCD) \\
    & & $12.17 \pm 1.35 \%$ ~\cite{SU3_Xic2Xlnu_2021} (SU(3)) \\
    & & $(11.9 \pm 1.3, 9.8 \pm 1.1, 10.7 \pm 0.9) \%$ ~\cite{SU3_Xic2Xlnu_Geng2019} (SU(3)) \\
    & & $11.3 \pm 3.4 \%$ ~\cite{LightFront_CharmBaryonSemi} (LFQM) \\
    & & $10.2 \pm 2.2 \%$ ~\cite{qcdsum_Xic2Xilnu} (QCDSR) \\
    & & $5.20 \pm 1.02 \%$ ~\cite{LightFront_Xic2Xienu} (LFQM) \\
    & & $5.39\%$ ~\cite{Lightfront_Xic2Xilnu_Zhao} (LFQM) \\
    & & $5.51 \pm 1.65 \%$ ~\cite{LCSR_Xic2Xilnu_2021} (LCSR) \\
    & & $9.40\%$ ~\cite{relativistitc_Xic2Xilnu} (RQM) \\
    \hline
    $\mathcal{B}(\Omega^0_c \to \Omega^- e^+ \nu_e )/\mathcal{B}(\Omega^0_c \to \Omega^- \pi^+ )$ & \multirow{2}{*}{CLEO II} & $1.1\pm 0.2$~\cite{LightFront_Qarkmodel_Omegac} (LFQM)\\
    =$2.4 \pm 1.2 $~\cite{PDG,CLEO_Omegac_Omegaenu} &  &  $0.71$~\cite{lcsr_Omegac_Omegalnu} (LCSR) \\
    \hline
    $\mathcal{B}(\Omega^0_c \to \Omega^- e^+ \nu_e )/\mathcal{B}(\Omega^0_c \to \Omega^- \pi^+ )$& \multirow{2}{*}{Belle} & \multirow{2}{*}{$0.9 \pm 0.1$~\cite{LightFront_Qarkmodel_Omegac} (LFQM), $0.71$~\cite{lcsr_Omegac_Omegalnu} (LCSR)}  \\
    =$1.98 \pm 0.13_\mathrm{stat} \pm 0.08_\mathrm{syst} $~\cite{Belle_Omegac_Omegalnu} & & \\
    $\mathcal{B}(\Omega^0_c \to \Omega^- \mu^+ \nu_\mu )/\mathcal{B}(\Omega^0_c \to \Omega^- \pi^+ )$ & \multirow{2}{*}{Belle} & \multirow{2}{*}{$0.9 \pm 0.1$~\cite{LightFront_Qarkmodel_Omegac} (LFQM), $0.68$~\cite{lcsr_Omegac_Omegalnu} (LCSR)}  \\
    =$1.94 \pm 0.18_\mathrm{stat} \pm 0.10_\mathrm{syst} $~\cite{Belle_Omegac_Omegalnu} & &  \\
    $\mathcal{B}(\Omega^0_c \to \Omega^- e^+ \nu_e)/\mathcal{B}(\Omega^0_c \to \Omega^- \mu^+ \nu_\mu) $& \multirow{2}{*}{Belle} & \multirow{2}{*}{$\approx 1.0$~\cite{Angular_LFU_CharmBaryon}} \\
    =$1.02\pm 0.10_\mathrm{stat} \pm 0.02_\mathrm{syst}$~\cite{Belle_Omegac_Omegalnu}&  & \\
    \hline\hline
  \end{tabular}
\end{table}

\subsection{Discussion: $|V_{cd}|$ and $|V_{cs}|$ determinations}
\label{sec:discussionVcdVcs}
\begin{figure}[hbpt]
\centering
\subfigure[]{\includegraphics[width=3.0in]{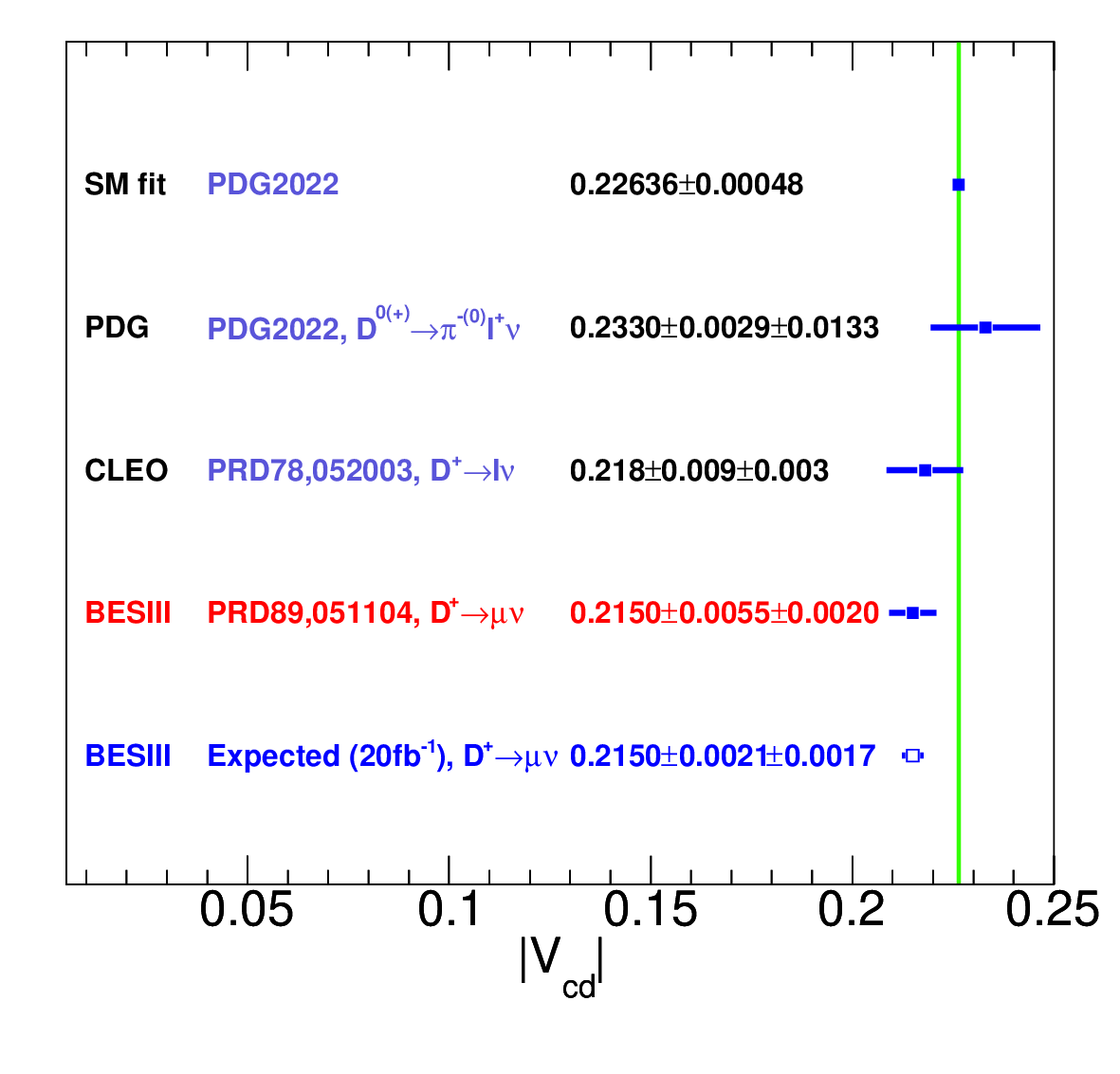}}\hskip -1pt
\subfigure[]{\includegraphics[width=3.0in]{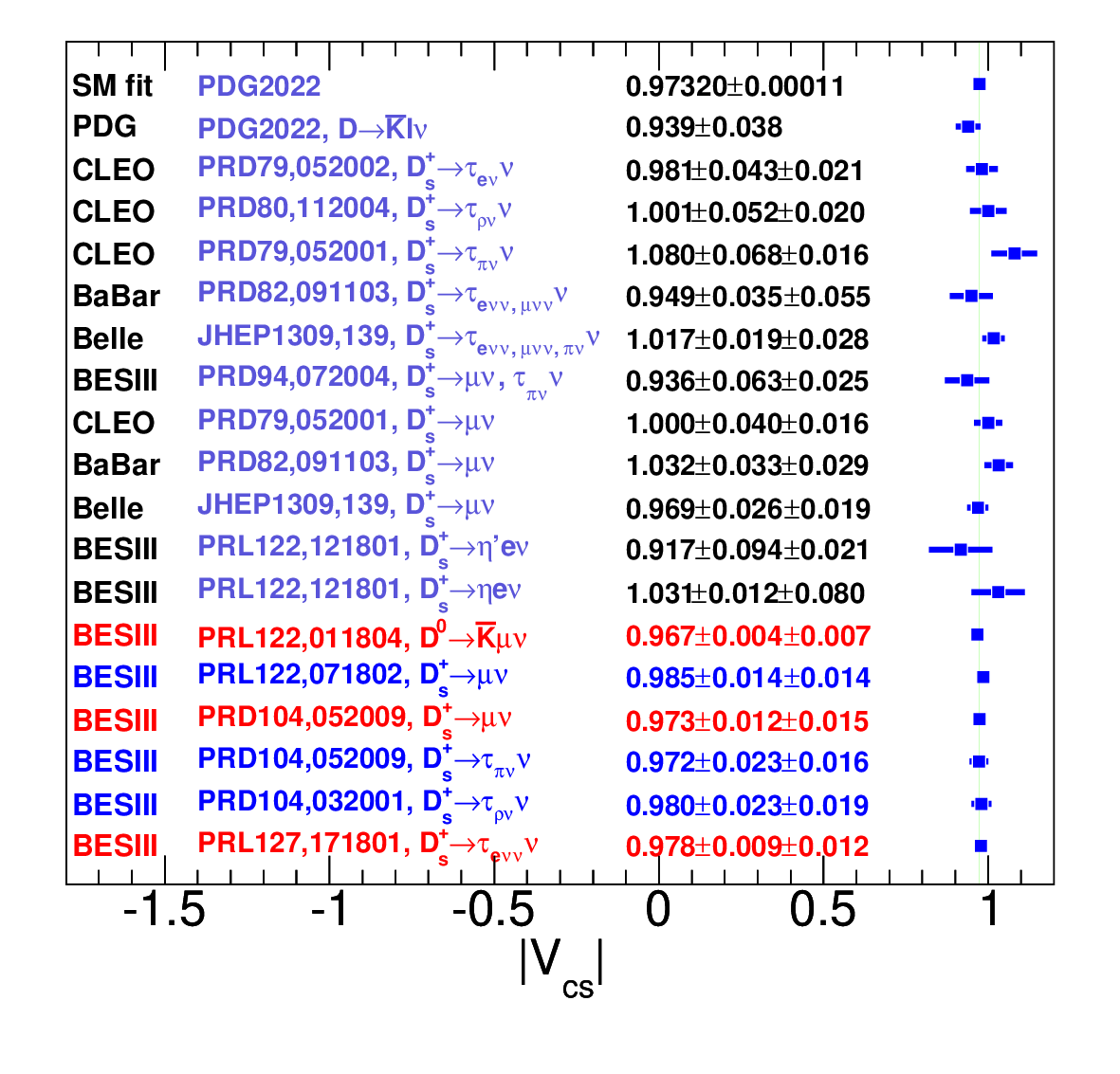}}
\caption{Precision of the measurements of (a) $|V_{cd}|$ and (b) $|V_{cs}|$. The green bands indicate the uncertainties of the average values from the global fit in SM~\cite{PDG}. (In the case of $|V_{cs}|$, the green error band is barely visible, as the SM unitarity constraint leads to very small uncertainty.) 
  The value marked in open square
  denotes the expected precision with the BESIII data sets that will be accumulated in the future~\cite{Ablikim:2019hff}.
  Data from Refs.~\cite{CLEO:2008ffk, bes3_muv,  CLEO:2009jky, CLEO:2009vke, CLEO:2009lvj, BaBar:2010ixw, Belle:2013isi, BESIII:2016cws, bes3_Dst_etaenu, bes3_D0_kmuv, bes3_Ds_muv, bes3_Ds_tauv1, bes3_Ds_tauv2, bes3_Ds_tauv3}.
  Figure provided by Hai-Long Ma}
\label{fig:ckm}
\end{figure}

The LQCD theory has made an admirable achievement in the past decade, improving
the uncertainties of calculated decay constants from the level of 1-2\% to
0.33\% ($f_D$) and 0.2\% ($f_{D_s}$)~\cite{FlavourLatticeAveragingGroupFLAG:2021npn}.
The experiments have not quite been able to keep up the pace:
the current PDG world average values of $|V_{cd}|$ and $|V_{cs}|$ struggle with
uncertainties of 2.5\% and 1.5\%, respectively, which are dominated by the 
BESIII leptonic-decay measurements. In these measurements, the systematic 
uncertainty on $|V_{cd}|$ is smaller than its
statistical uncertainty, while the statistical and systematic uncertainties of
$|V_{cs}|$ are comparable, as shown in Fig.~\ref{fig:ckm}.

In addition, the measurements of the charmed-meson semi-leptonic decays
$D^{0(+)} \to \bar{K} \ell \nu_\ell$ and $D^{0(+)} \to \pi \ell \nu_\ell$
with form factors from LQCD calculations as the key inputs~\cite{FlavourLatticeAveragingGroupFLAG:2021npn}
offer a completely different way of extracting $|V_{cs}|$ and
$|V_{cd}|$. The experimental data and the latest $n_f=2+1+1$ LQCD results on $D\to \bar{K}\ell\nu_\ell$ 
decay are so precise, that $|V_{cs}|$ can be determined with $\le 1$\% accuracy.
In~\cite{Chakraborty:2021qav} the authors tested three different ways of determining $|V_{cd}|$
from their LQCD form factors and experimental data: 1) using the full kinematic $q^2$ range and
differential branching fractions $\mathrm{d}\Gamma/\mathrm{d}q^2$; 2) using their determination
of the value of the form factor at $q^2=0$, $f_+(0)$, and experimental determinations of $|V_{cs}|f_+(0)$;
3) Using the integrated, total brancing fraction $\mathcal{B}$, integrating the LQCD form factor over
the whole $q^2$ range. All three methods gave results that are within $0.6\sigma$, and with similar 
precision.
However, the lattice results for $D\to \pi\ell\nu_\ell$ have much larger uncertainties.
In this review we have opted to use the $N_f=2+1+1$ FLAG results~\cite{FlavourLatticeAveragingGroupFLAG:2021npn}
for both $D\to \bar{K}\ell\nu_\ell$ and $D\to\pi\ell\nu_\ell$ decays for consistency, even though the $n_f=2+1$ lattice
results for the latter decay have 4.4\% uncertainty compared to 5.7\% for the $n_f=2+1+1$ results.
The situation should improve in the next year or so, as two lattice collaborations are working
on extracting the $D\to\pi\ell\nu_\ell$ form factors with better precision~\cite{FermilabLattice:2021bxu,Marshall:2022xbz}. 
The anticipated uncertainties should allow the extraction of $V_{cd}$ from semileptonic decays to precision of
$\sim 1$\%.

\begin{figure}[hbpt]
\centering
\includegraphics[width=0.5\linewidth]{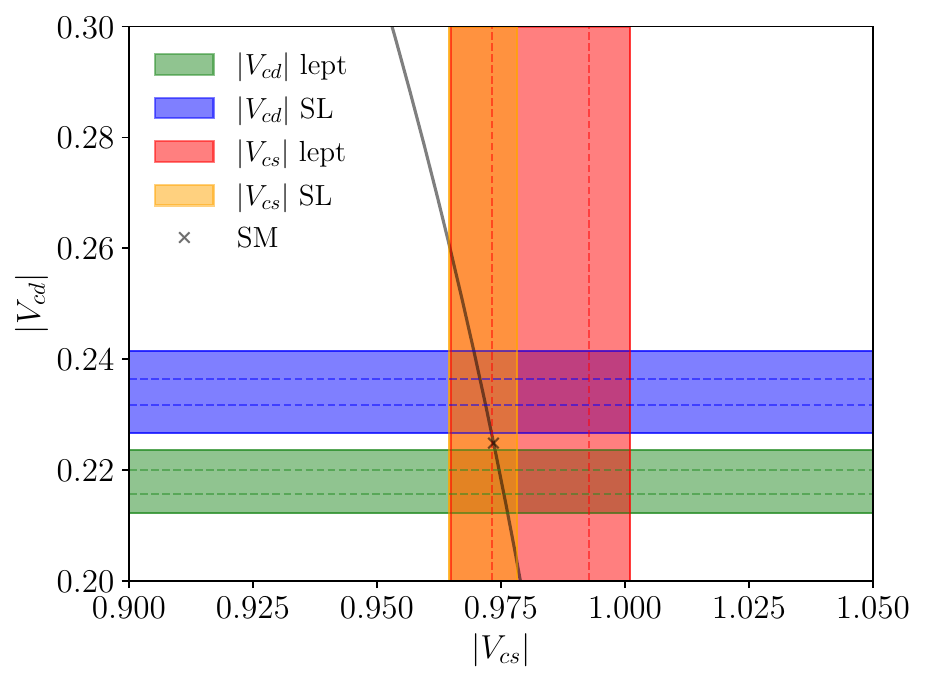}
  \caption{Summary of the results from leptonic and semileptonic (SL) decays in the $|V_{cs}|$--$|V_{cd}|$
  plane (see also Eq.~\eqref{eq:lept_SL_results}). The leptonic and semileptonic determinations are the
$n_f=2+1+1$ results from the FLAG 2021 review~\cite{FlavourLatticeAveragingGroupFLAG:2021npn}, and the cross
labelled SM is the result from SM fits assuming unitarity from PDG~\cite{PDG}. The black line shows the
unitarity constraint $|V_{cd}|^2+|V_{cs}|^2=1-|V_{cb}|^2$ using 
$|V_{cb}| = (41.0\pm 1.4)\times 10^{-3}$~\cite{PDG}. The uncertainty on $|V_{cb}|$ is smaller than the width of 
the line. The dashed lines show the anticipated 1\% uncertainties in the future.}
\label{fig:Vcd_vs_Vcs}
\end{figure}

Let us now collect and compare the results from leptonic and semileptonic
decays for the CKM elements:
\begin{align}
\label{eq:lept_SL_results}
|&V_{cd}|^{\textrm{lept}}= 0.2179\pm57, &&|V_{cs}|^{\textrm{lept}}= 0.983\pm18\nonumber \\
&|V_{cd}|^{\textrm{SL}}= 0.2341\pm74, &&|V_{cs}|^{\textrm{SL}}= 0.9714\pm69\\
&|V_{cd}|^{\textrm{SM}}=0.22486\pm67  && |V_{cs}|^{\textrm{SM}}=0.97349\pm16.\nonumber 
\end{align}
The first two rows, the leptonic and semileptonic (SL) determinations, are the
$n_f=2+1+1$ results from the FLAG 2021 review~\cite{FlavourLatticeAveragingGroupFLAG:2021npn}.
The third row gives the SM fits assuming unitarity from PDG~\cite{PDG}. 
These results are also illustrated in Fig.~\ref{fig:Vcd_vs_Vcs}.
There are
some tensions between the leptonic and semileptonic determinations, as well as with the fits
assuming unitarity, at the $1.7\sigma$ level. However, the uncertainties from the leptonic
and semileptonic determinations are much larger than the ones from the global fits that impose the
unitarity constraint. The test of the second row unitarity is limited by the precision of $|V_{cs}|$.
Treating the values as uncorrelated and using $|V_{cb}| = (41.0\pm 1.4)\times 10^{-3}$~\cite{PDG}, 
one gets $|V_{cd}|^2+|V_{cs}|^2+|V_{cb}|^2=1.015\pm 0.035$ for the leptonic and
$|V_{cd}|^2+|V_{cs}|^2+|V_{cb}|^2=1.000\pm14$ for the semileptonic determinations. 
For comparison, the PDG~\cite{PDG} report $|V_{ud}|^2+|V_{us}|^2+|V_{ub}|^2 = 0.9985\pm 0.0005$ for the test of 
unitarity of the first row, where the uncertainties are much smaller.

One can also calculate the ratio
$|V_{cd}/V_{cs}|^2 = 0.048\pm 0.003_\mathrm{stat}\pm 0.001_\mathrm{syst}$,
with the BESIII results and the FLAG~\cite{FlavourLatticeAveragingGroupFLAG:2021npn}
value for
$(f_{D^+_s}/f_{D^+})^\mathrm{FLAG} = 1.1783 \pm 0.0016$. This value deviates
from the expectation given by the CKMfitter group~\cite{CKMFitter}, 
$|V_{cd}/V_{cs}|^2 = 0.05335\pm 0.00011$, by 1.7$\sigma$. 

The uncertainties of $|V_{cd}|$ and $|V_{cs}|$ measurements are currently
dominated by the limited experimental statistics. With the $\sim20$~fb$^{-1}$
$\psi(3770)$ data being taken by BESIII~\cite{Ablikim:2019hff}, it can be
expected that the statistical uncertainty will be comparable to the 
uncertainty arising from the LQCD calculations of decay constants. The
relative uncertainty on the $|V_{cs}|$ and $|V_{cd}|$ determinations will both
be improved to the 1\% level. If the $|V_{cd}|$ result is the same as its
current measured value, the significance of the discrepancy would increase to
about the 4$\sigma$ level as shown in Figs.~\ref{fig:ckm}(a) and~\ref{fig:Vcd_vs_Vcs}.

\subsection{Lepton flavor universality} 
LFU is one of key predictions in the SM and
indicates that three generations of leptons share equal coupling to gauge
bosons. Specifically, the purely leptonic decay widths of charmed mesons,
Eq.~\eqref{eq01}, is proportional to the lepton mass-squared, which is the
consequence of the helicity suppression. Their ratios between different 
leptons ($e^+\nu_e:\mu^+\nu_\mu:\tau^+\nu_\tau$) depend only on the lepton 
masses and can be accurately predicted to be $2.35\times10^{-5}:1:2.67$ for 
$D^+$ and $2.35\times10^{-5}:1:9.75$ for $D_s^+$ based on SM with negligible 
uncertainty (the only uncertainty is coming from the determination of the 
lepton masses). The $D_{(s)}^+\to e^+\nu_e$ decay, with 
a $<10^{-8}$
expected branching fraction, is not experimentally observed yet, but comparing
the obtained branching fractions of $D_{(s)}^+ \to \tau^+\nu_\tau$ and
$D_{(s)}^+ \to \mu^+\nu_\mu$ gives important comprehensive test of $\tau\textrm{-}\mu$
lepton-flavor universality. Using the world average PDG values~\cite{PDG}, one can determine 
$\Gamma(D^{+}\to\tau^+\nu_\tau)/\Gamma(D^{+}\to\mu^+\nu_\mu)=3.21\pm0.74$ and $\Gamma(D_{s}^{+}\to\tau^+\nu_\tau)/\Gamma(D_{s}^{+}\to\mu^+\nu_\mu)=9.80\pm0.34$
which, although still statistically limited, are consistent with the SM 
predictions.

\begin{figure}[hbpt]
  \centering
 \includegraphics[width=0.48\textwidth]{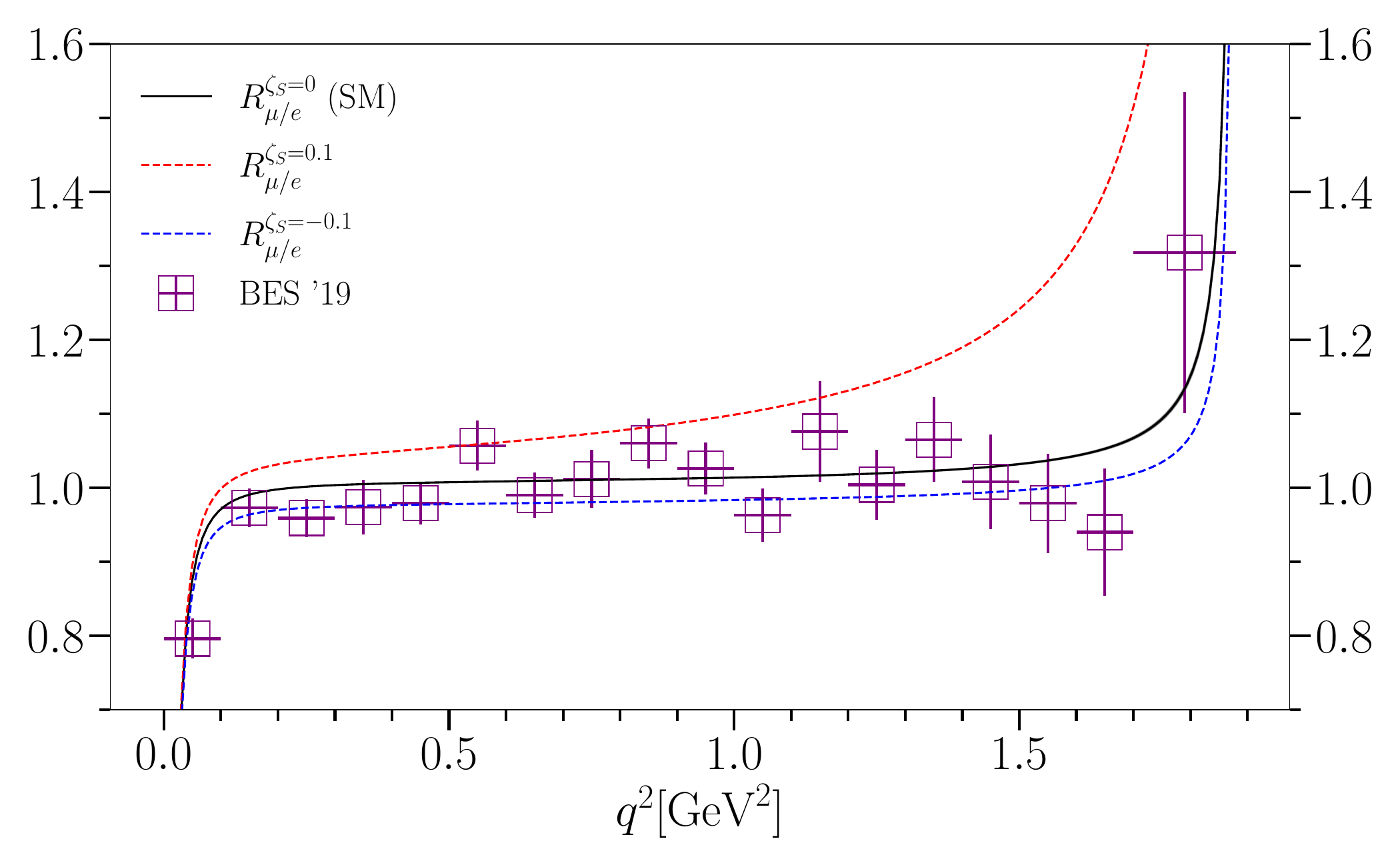}
\caption{Lepton flavour universality tests in $D \to \bar{K}\ell\nu_\ell$ decay. The solid black
    curve as a function of $q^2$ shows the SM ratio of branching fractions
    for a muon in the final state to that for an electron obtained from LQCD form factors
    (for details see Ref.~\cite{Chakraborty:2021qav} where this plot is from). The width of 
    the curve gives the (very small) uncertainty from LQCD. Possible QED effects are not 
    included. The points, with error bars, are from the BESIII experiment~\cite{bes3_D0_kmuv}. 
    The red and blue dashed lines show what the curve would look like in the presence of a
    new physics scalar coupling for the $\mu$ case (for details see
    Ref.~\cite{Chakraborty:2021qav}). Figure from Ref.~\cite{Chakraborty:2021qav}.}
\label{fig:LFU}
\end{figure}

Analogously, the ratios of electronic and muonic semileptonic decay widths of 
charmed mesons
can also be accurately calculated based on Eq.~\eqref{eq:semi}, while 
decays involving $\omega, \eta, \rho$ receive about 5\% uncertainty from the 
form factor models. These predictions along with the experimental results 
examine the LFU in the $e\textrm{-}\mu$ sector. However, suffering from experimental 
uncertainty of several percent to few ten percent, no significant LFU 
violation has been observed yet in 
the charm sector. The best test of $\mu\textrm{-}e$ LFU for semi-leptonic 
$D^{0(+)}_{(s)}$ decays is expected to be from $D\to \bar{K}\ell^+\nu_\ell$ 
decays, where the precision of the test can be reduced from 1.3\% to the level of 
0.8\% with the forthcoming BESIII data. Comparison of the SM prediction
of the $\mu$-$e$ LFU as a function of $q^2$ in the $D\to \bar{K}\ell\nu_\ell$ decay
calculated using LQCD form factors and the current experimental 
results from BESIII~\cite{bes3_D0_kmuv} is shown in Fig.~\ref{fig:LFU} (taken
from \cite{Chakraborty:2021qav}). The figure highlights the fact that the uncertainties
from LQCD are negligible compared to experimental uncertainties, and the
aforementioned increase in statistics is needed to detect any possible violations
in LFU in $D\to \bar{K}\ell\nu_\ell$, $\ell=e,\mu$ decays.
At present, it is not conclusive about 
whether the $\mu$-$e$ LFU always holds in semi-leptonic $D^{0(+)}_{(s)}$ decays, 
as there are still many un-observed semi-muonic decays, 
e.g.~$D^+\to \eta^\prime \mu^+\nu_\mu$, $D^{0(+)}\to a_0(980) \mu^+\nu_\mu$,
$D^{0(+)}\to K_1(1270) \mu^+\nu_\mu$, $D^+\to f_0(500) \mu^+\nu_\mu$, 
$D^+_s\to K^0\mu^+\nu_\mu$, $D^+_s\to K^{*0}\mu^+\nu_\mu$,
$D^+_s\to f_0(980)\mu^+\nu_\mu$, and $D^+_s\to \eta^\prime \mu^+\nu_\mu$.
Larger data samples will offer opportunities to search for these decays and
thereby clarify if there is possible violation of $\mu$-$e$ LFU in the charmed 
meson sector. For the semi-leptonic charmed baryon decays, there are 
experimental results of $\Lambda_c^{+}\to\Lambda \ell^+\nu_\ell$ by 
BESIII~\cite{2022_Lamdbac_Lenu, Ablikim:2016vqd}(Table~\ref{tab:semi}),
$\Omega^0_c \to \Omega^- \ell^+ \nu_\ell$ and $\Xi^0_c \to  \Xi^- \ell^+ \nu_\ell$ by 
Belle~\cite{belle_Xic_Xilnu, Belle_Omegac_Omegalnu}(Table~\ref{tab:Xic_semi}), 
which test the $\mu$-$e$ LFU in the charmed baryon sector.

\subsection{Rare leptonic and semileptonic $D_{(s)}$ decays} 
The rare leptonic and semileptonic decays of the $D_{(s)}$ meson provide a 
sensitive probe 
for new physics. In recent years, the Belle, BaBar, LHCb, and BESIII 
experiments performed many new measurements to search for new physics in this 
area.

Charm decays with two oppositely charged leptons ($\ell^+\ell^-$) in the final 
state may proceed via the quark flavor-changing neutral-current (FCNC) process 
$c \to u\ell^+\ell^-$. FCNC processes are forbidden at tree level in the SM and 
only allowed in loop and box diagrams, which are highly suppressed by the 
Glashow-Iliopoulos-Maiani (GIM) mechanism. Therefore, they provide a unique 
opportunity to search contributions from new physics beyond the SM via couplings 
to up-type quarks. The expected branching fractions of the FCNC decays 
$D^0\to e^+e^-$ and $D^0\to \mu^+\mu^-$ in the SM are at $\mathcal{O}(10^{-13})$.
Nowadays the sensitivity of experimental measurements for the branching fractions 
has reach $10^{-9}$~\cite{BaBar:2004uxd, Belle_PureLep_2010, BaBar:2012ujp, LHCb:2013jyo}, 
but no evidence is observed yet. The GIM suppression is more effective for the 
charm sector compared to the down-type quarks in the bottom and strange sectors. 
This suppression is responsible for the relatively small size of charm mixing and 
CP violation in the charm system. Therefore, FCNC processes in 
$D\to h\ell^+\ell^-$ decays are often totally overshadowed by long distance 
contributions, where $h$ denotes one or more hadrons. The short-distance 
contributions from FCNC to the branching fraction of $D\to h\ell^+\ell^-$ are 
expected to the order of $\mathcal{O}(10^{-9})$, while long-distance contributions from 
tree-level processes involving intermediate vector resonances, 
i.e.~$D\to hV(\to\ell^+\ell^-)$, are predicted to increase the branching 
fraction to $\mathcal{O}(10^{-6})$. Many relevant attempts have been made to 
impose constraints on physics beyond the SM~\cite{BaBar:2011ouc, LHCb:2013hxr, LHCb:2017yqf, JHEP.2021.044, PhysRevD.105.L071102, LHCb:2013gqe, 2016558, LHCb:2017uns, LHCb:2018qsd, PhysRevLett.122.081802, BaBar:2019hxw, PhysRevLett.128.221801}.

Decay modes with two oppositely charged leptons of different flavor correspond 
to lepton flavor violating (LFV) decays and are essentially forbidden in the SM 
because they can occur only through lepton mixing. Decays involving two 
same-sign leptons are both LFV and lepton-number violating (LNV) decays, which 
is also strictly suppressed in the SM. An interesting source of LNV processes 
is given by exchanging a single Majorana neutrino with a mass on the order of 
the heavy flavor mass scale, where the Majorana neutrino can be kinematically 
accessible and produced on shell. Although no significant signal is observed so 
far, the experimental results provide the supplementary information in the 
study of lepton-mixing and the nature of neutrinos~\cite{Belle_PureLep_2010, BaBar:2011ouc, BaBar:2012ujp, LHCb:2015pce, BaBar:2019hxw, BaBar:2020faa, JHEP.2021.044, BESIII:2019oef, BaBar:2019hxw, JHEP.2021.044}.

\section{Discussion and outlook}
The SM is the greatest successful theoretical framework of particle 
physics, but there are still a number of issues that deserve further both 
experimental and theoretical investigation. Purely leptonic and semileptonic 
decays of charmed hadrons remain the best platform to test the SM 
in the charm sector. In this review, we have summarized the recent results obtained 
in the BESIII experiments with data sets collected 
at the production thresholds
of $D\bar{D}$, $D_s^{*+}D_s^{-}$ and $\Lambda_c^{+}\bar{\Lambda}_c^{-}$, and 
review the theoretical and experimental tools used. 
We have also summarized the updated measurements of semileptonic $\Xi_c^{0,+}$ and 
$\Omega_c$ decays by Belle, CLEO II, and ALICE experiments.
These unique opportunities provide rigorous tests of QCD-based models,
the unitarity of CKM matrix, and leptonic-flavor universality in the charm
energy region. 

Our knowledge of hadronic charm purely leptonic and semileptonic decays has 
improved significantly over the last few years. The values of the decay
constants and form factors can be calculated in high precision from first principles using
LQCD. The obtained precision for 
the $f_{D^{+}_{(s)}}$ calculations has reached the $0.2$--$0.3$\% level, and can 
translate leptonic decay rate measurements into high precision determinations 
of the CKM matrix elements $|V_{cd}|$ and $|V_{cs}|$. Moreover, the highly 
accurate prediction 
of $f_{D^{+}_{(s)}}$ and form factors by LQCD allow detailed theoretical studies 
of the charmed hadron decay dynamics, such as SU(3). 
However, the current experimental results can only provide relative low-precision
calibration or tests of theoretical model calculations with limited amount of data samples.

The data samples at the  $D\bar{D}$ and $D_s^{*+}D_s^{-}$ thresholds to be 
collected by BESIII in the coming years offer opportunities to further improve 
the precision of the measurements of these important constants. The relative uncertainties on 
$|V_{cd}|f_{D^+}$ and $|V_{cs}|f_{D^+_s}$ can be reduced from 2.6\% 
and 1.2\% to approximately 
1.1\% and 0.9\%, respectively, and the systematic uncertainty is expected to 
dominate at that time. 
As the leptonic determinations of the CKM matrix elements have uncertainties that are
reaching the percent level, higher-order electroweak and hadronic-structure
dependent corrections to the decay rate become important. These have not been computed for 
$D_{(s)}$ meson decays yet, but for pion and kaon decays they have 
been estimated to be around 1–2\% \cite{Cirigliano:2007ga}.
It is therefore important that such theoretical calculations are tackled in the
near future.
The uncertainties of lepton-flavor universality tests in $D^+ \to \ell^+\nu_\ell$ and 
$D_{s}^+ \to \ell^+\nu_\ell$ decays are also expected to be reduced from 24.0\% 
and 4.0\% to about 10.0\% and 3.0\%, respectively.

The precision of all measurements of semi-leptonic $D_{(s)}^{0(+)}\to P$ and 
$D_{(s)}^{0(+)}\to V$ form factors, except for the $D^{0(+)}\to K$ and 
$D^{0(+)}\to K^*$, are restricted due to limited data sets.
With BESIII future data samples, all the form-factor measurements which are 
currently statistically limited will be statistically improved by a factor of 
up to 2.6 and 1.4 for semi-leptonic $D^{0(+)}$ snd $D^+_s$ decays, 
respectively.
The dynamics 
studies of the semi-leptonic $D\to P$ decays will offer complementary 
measurements of $|V_{cs}|$ and $|V_{cd}|$ using the whole kinematic $q^2$ range. 
The current precisions of the $|V_{cd}|$ measurements with semileptonic $D^{0(+)}$ decays 
are limited by the theoretical uncertainties of the form factors in 
LQCD, with FLAG quoting a 4.4\% uncertainty for $f^{D\to \pi}_+(0)$ from $n_f=2+1$ calculations
and 5.7\% from $n_f=2+1+1$ calculations~\cite{FlavourLatticeAveragingGroupFLAG:2021npn}. 
Here the situation can be expected to improve within the next year or two, as two collaborations
are finalising their ongoing calculations~\cite{Marshall:2022xbz,FermilabLattice:2021bxu}
\footnote{We note that a new
paper by Fermilab Lattice and MILC collaborations has appeared in the arXiv~\cite{FermilabLattice:2022gku}, after
this review article had already been submitted to ARNP.},
and the determination of $V_{cd}$ from semileptonic decays with $\sim 1$\% uncertainties should become possible.
For $|V_{cs}|$ the situation is much better, as lattice calculations have reached below 1\%
precision for the form factors of $D\to \bar{K}\ell\nu_\ell$ decays,
thus allowing the extraction of $|V_{cs}|$ at similar precision (FLAG~\cite{FlavourLatticeAveragingGroupFLAG:2021npn}
quote a 0.7\% uncertainty from their fit). At his precision, incorporation of
electromagnetic corrections from first principles is a necessary step in the near future.
In addition, the forthcoming BESIII data will allow to extract the 
$D\to S$ and $D\to A$ form factors from experiment for the first time.

Charmed hadron studies will continue during the future upgrade of the BESIII 
experiment. BESIII plans to collect 7 times the current amount of $D\bar{D}$
thereshold data, and this will usher in a precision charm flavor era.
Along with the improvements in the LQCD calculations on the decay constants and form
factors expected circa 2025, we can then anticipate significantly improved 
constraints on the ( $|V_{cs}|$, $|V_{cd}|$) plane~\cite{Ablikim:2019hff}
This will allow for precise tests of the 
consistency of CKM determinations from different quark 
sectors~\cite{Ablikim:2019hff, Charles:2004jd}.

Following BESIII, the Super Tau-Charm Facility~(STCF)~\cite{STCF_LoI} has been proposed in China, which 
is a symmetric electron-positron collider operating at $\sqrt{s}$ from 2.0 to 7.0~GeV. The energy region of STCF covers the pair production 
thresholds for $D^{0(+)}$, $D_s^+$, $\Lambda_c^{+}$, $\Xi_c^{0(+)}$ and 
$\Omega_c^{0}$ hadrons. The peak luminosity is designed to be over 
$0.5\times10^{35}$ cm$^{-2}$s$^{-1}$ at $4.0$~GeV and the designed 
integrated luminosity per year is approximate 1 ab$^{-1}$, corresponding to a 
data rate about a factor 100 larger than the present BEPCII.
The STCF will become a precision frontier for
exploring the nature of non-perturbative strong interactions, understanding the
internal structure of hadrons, studying the CP violation of hadron decays and
searching for the asymmetry of matter-antimatter, and testing the SM and probing
for physics beyond the SM with unprecedented sensitivity.


\section*{ACKNOWLEDGMENTS}
This work is supported in part by National Key R\&D Program of China under Contracts Nos. 2020YFA0406300, 2020YFA0406400; National Natural Science Foundation of China (NSFC) under Contracts Nos. 11935018, 11875054, 12192263, 11935015, 12221005; the Chinese Academy of Sciences (CAS) Large-Scale Scientific Facility Program; Joint Large-Scale Scientific Facility Funds of the NSFC and CAS under Contracts No. U2032104;
J.K. acknowledges support by the European Research Council (ERC) under the European Union’s Horizon 2020 research and innovation program through Grant Agreement No. 771971-SIMDAMA.

\bibliography{review_bib}{}
\bibliographystyle{unsrt}
\end{document}